\documentclass[10pt, journal, final]{IEEEtran}
\usepackage[draft]{hyperref}  
\usepackage{gensymb}
\usepackage{dsfont}
\usepackage{amsmath,amssymb,amsfonts,amsthm}
\DeclareMathOperator{\sinc}{sinc}
\usepackage{epsfig}
\usepackage{cite}
\usepackage{hhline}
\usepackage{multirow}
\usepackage{xcolor}
\usepackage{makecell}
\usepackage{subcaption}
\usepackage{capt-of}
\usepackage[labelformat=simple]{subcaption}

\usepackage{enumerate}
\usepackage{enumitem}
\usepackage{color}
\usepackage{graphicx}
\usepackage{algpseudocode}
\usepackage[ruled]{algorithm}
\makeatletter
\newcounter{parentalgorithm}

\makeatother

\usepackage[nolist,printonlyused]{acronym}      
\usepackage{booktabs}
\usepackage{bm}
\usepackage{pgf}
\usepackage{pgfplots}
\pgfplotsset{
        layers/my layer set/.define layer set={
            background,
            main,
            foreground
        }{
        },
set layers=my layer set,
}
\usepackage{tikz}
\usepackage{grffile}
\pgfplotsset{compat=newest}
\usetikzlibrary{plotmarks}
\usetikzlibrary{quotes,angles,automata,arrows,positioning,calc,3d,matrix,decorations.markings,matrix}

\makeatletter
\newcommand{\vast}{\bBigg@{3}}
\newcommand{\Vast}{\bBigg@{4}}
\makeatother
\renewcommand{\IEEEQED}{\IEEEQEDopen}

\showoutput
\makeatletter
\tikzset{%
  remember picture with id/.style={%
    remember picture,
    overlay,
    save picture id=#1,
  },
  save picture id/.code={%
    \edef\pgf@temp{#1}%
    \immediate\write\pgfutil@auxout{%
      \noexpand\savepointas{\pgf@temp}{\pgfpictureid}}%
  },
  if picture id/.code args={#1#2#3}{%
    \@ifundefined{save@pt@#1}{%
      \pgfkeysalso{#3}%
    }{
      \pgfkeysalso{#2}%
    }
  }
}

\def\savepointas#1#2{%
  \expandafter\gdef\csname save@pt@#1\endcsname{#2}%
}

\def\tmk@labeldef#1,#2\@nil{%
  \def\tmk@label{#1}%
  \def\tmk@def{#2}%
}

\tikzdeclarecoordinatesystem{pic}{%
  \pgfutil@in@,{#1}%
  \ifpgfutil@in@%
    \tmk@labeldef#1\@nil
  \else
    \tmk@labeldef#1,(0pt,0pt)\@nil
  \fi
  \@ifundefined{save@pt@\tmk@label}{%
    \tikz@scan@one@point\pgfutil@firstofone\tmk@def
  }{%
  \pgfsys@getposition{\csname save@pt@\tmk@label\endcsname}\save@orig@pic%
  \pgfsys@getposition{\pgfpictureid}\save@this@pic%
  \pgf@process{\pgfpointorigin\save@this@pic}%
  \pgf@xa=\pgf@x
  \pgf@ya=\pgf@y
  \pgf@process{\pgfpointorigin\save@orig@pic}%
  \advance\pgf@x by -\pgf@xa
  \advance\pgf@y by -\pgf@ya
  }%
}

\makeatother

\usepackage{comment}
\usepackage{xpatch}

\tikzset{
    ncbar angle/.initial=90,
    ncbar/.style={
        to path=(\tikztostart)
        -- ($(\tikztostart)!#1!\pgfkeysvalueof{/tikz/ncbar angle}:(\tikztotarget)$)
        -- ($(\tikztotarget)!($(\tikztostart)!#1!\pgfkeysvalueof{/tikz/ncbar angle}:(\tikztotarget)$)!\pgfkeysvalueof{/tikz/ncbar angle}:(\tikztostart)$)
        -- (\tikztotarget)
    },
    ncbar/.default=0.5cm,
}

\begin{document}

\title{A Hybrid Model-based and Data-driven Approach to Spectrum Sharing in mmWave Cellular Networks}

\author{Hossein S. Ghadikolaei, Hadi Ghauch,~\IEEEmembership{Member,~IEEE,}
        Gabor~Fodor,~\IEEEmembership{Senior~Member,~IEEE,}\\
    Mikael~Skoglund,~\IEEEmembership{Fellow, IEEE,}~and
        Carlo~Fischione,~\IEEEmembership{Senior Member,~IEEE}

\thanks{H. S. Ghadikolaei, G. Fodor, M. Skoglund, and C. Fischione are with the School of Electrical Engineering and Computer Science, KTH Royal Institute of Technology, 100 44 Stockholm, Sweden (e-mail: \{hshokri, gaborf, skoglund, carlofi\}@kth.se).}
\thanks{H. Ghauch is with the COMELEC Department, Telecom ParisTech, France (e-mail: {hadi.ghauch}@telecom-paristech.fr).}
\thanks{This work was partially sponsored by the Ericsson project SPECS~II and the Swedish Research Council under grant 2018-00820.}
}


\newtheorem{theorem}{Theorem}
\newtheorem{defin}{Definition}
\newtheorem{prop}{Proposition}
\newtheorem{lemma}{Lemma}
\newtheorem{corollary}{Corollary}
\newtheorem{alg}{Algorithm}
\newtheorem{remark}{Remark}
\newtheorem{result}{Result}
\newtheorem{conjecture}{Conjecture}
\newtheorem{example}{Example}
\newtheorem{notations}{Notations}
\newtheorem{assumption}{Assumption}

\newcommand{\combin}[2]{\ensuremath{ \left( \ba{c} #1 \\ #2 \ea \right) }}
\newcommand{\diag}{{\mbox{diag}}}
\newcommand{\rank}{{\mbox{rank}}}
\newcommand{\dom}{{\mbox{dom{\color{white!100!black}.}}}}
\newcommand{\range}{{\mbox{range{\color{white!100!black}.}}}}
\newcommand{\image}{{\mbox{image{\color{white!100!black}.}}}}
\newcommand{\herm}{^{\mbox{\scriptsize H}}}  
\newcommand{\sherm}{^{\mbox{\tiny H}}}       
\newcommand{\tran}{^{\mbox{\scriptsize T}}}  
\newcommand{\tranIn}{^{\mbox{-\scriptsize T}}}  
\newcommand{\card}{{\mbox{\textbf{card}}}}
\newcommand{\asign}{{\mbox{$\colon\hspace{-2mm}=\hspace{1mm}$}}}
\newcommand{\ssum}[1]{\mathop{ \textstyle{\sum}}_{#1}}

\newcommand{\vbar}{\raisebox{.17ex}{\rule{.04em}{1.35ex}}}
\newcommand{\vbarind}{\raisebox{.01ex}{\rule{.04em}{1.1ex}}}
\newcommand{\D}{\ifmmode {\rm I}\hspace{-.2em}{\rm D} \else ${\rm I}\hspace{-.2em}{\rm D}$ \fi}
\newcommand{\T}{\ifmmode {\rm I}\hspace{-.2em}{\rm T} \else ${\rm I}\hspace{-.2em}{\rm T}$ \fi}
\newcommand{\B}{\ifmmode {\rm I}\hspace{-.2em}{\rm B} \else \mbox{${\rm I}\hspace{-.2em}{\rm B}$} \fi}
\newcommand{\Hil}{\ifmmode {\rm I}\hspace{-.2em}{\rm H} \else \mbox{${\rm I}\hspace{-.2em}{\rm H}$} \fi}
\newcommand{\C}{\ifmmode \hspace{.2em}\vbar\hspace{-.31em}{\rm C} \else \mbox{$\hspace{.2em}\vbar\hspace{-.31em}{\rm C}$} \fi}
\newcommand{\Cind}{\ifmmode \hspace{.2em}\vbarind\hspace{-.25em}{\rm C} \else \mbox{$\hspace{.2em}\vbarind\hspace{-.25em}{\rm C}$} \fi}
\newcommand{\Q}{\ifmmode \hspace{.2em}\vbar\hspace{-.31em}{\rm Q} \else \mbox{$\hspace{.2em}\vbar\hspace{-.31em}{\rm Q}$} \fi}
\newcommand{\Z}{\ifmmode {\rm Z}\hspace{-.28em}{\rm Z} \else ${\rm Z}\hspace{-.38em}{\rm Z}$ \fi}

\newcommand{\sgn}{\mbox {sgn}}
\newcommand{\var}{\mbox {var}}
\newcommand{\E}{\mbox {E}}
\newcommand{\cov}{\mbox {cov}}
\renewcommand{\Re}{\mbox {Re}}
\renewcommand{\Im}{\mbox {Im}}
\newcommand{\cum}{\mbox {cum}}

\renewcommand{\vec}[1]{{\bf{#1}}}     

\newcommand{\vecsc}[1]{\mbox {\boldmath \scriptsize $#1$}}     
\newcommand{\itvec}[1]{\mbox {\boldmath $#1$}}
\newcommand{\itvecsc}[1]{\mbox {\boldmath $\scriptstyle #1$}}
\newcommand{\gvec}[1]{\mbox{\boldmath $#1$}}

\newcommand{\balpha}{\mbox {\boldmath $\alpha$}}
\newcommand{\bbeta}{\mbox {\boldmath $\beta$}}
\newcommand{\bgamma}{\mbox {\boldmath $\gamma$}}
\newcommand{\bdelta}{\mbox {\boldmath $\delta$}}
\newcommand{\bepsilon}{\mbox {\boldmath $\epsilon$}}
\newcommand{\bvarepsilon}{\mbox {\boldmath $\varepsilon$}}
\newcommand{\bzeta}{\mbox {\boldmath $\zeta$}}
\newcommand{\boldeta}{\mbox {\boldmath $\eta$}}
\newcommand{\btheta}{\mbox {\boldmath $\theta$}}
\newcommand{\bvartheta}{\mbox {\boldmath $\vartheta$}}
\newcommand{\biota}{\mbox {\boldmath $\iota$}}
\newcommand{\blambda}{\mbox {\boldmath $\lambda$}}
\newcommand{\bmu}{\mbox {\boldmath $\mu$}}
\newcommand{\bnu}{\mbox {\boldmath $\nu$}}
\newcommand{\bxi}{\mbox {\boldmath $\xi$}}
\newcommand{\bpi}{\mbox {\boldmath $\pi$}}
\newcommand{\bvarpi}{\mbox {\boldmath $\varpi$}}
\newcommand{\brho}{\mbox {\boldmath $\rho$}}
\newcommand{\bvarrho}{\mbox {\boldmath $\varrho$}}
\newcommand{\bsigma}{\mbox {\boldmath $\sigma$}}
\newcommand{\bvarsigma}{\mbox {\boldmath $\varsigma$}}
\newcommand{\btau}{\mbox {\boldmath $\tau$}}
\newcommand{\bupsilon}{\mbox {\boldmath $\upsilon$}}
\newcommand{\bphi}{\mbox {\boldmath $\phi$}}
\newcommand{\bvarphi}{\mbox {\boldmath $\varphi$}}
\newcommand{\bchi}{\mbox {\boldmath $\chi$}}
\newcommand{\bpsi}{\mbox {\boldmath $\psi$}}
\newcommand{\bomega}{\mbox {\boldmath $\omega$}}

\def\calA{{\mathcal A}}
\def\calB{{\mathcal B}}
\def\calC{{\mathcal C}}
\def\calD{{\mathcal D}}
\def\calE{{\mathcal E}}
\def\calF{{\mathcal F}}
\def\calG{{\mathcal G}}
\def\calH{{\mathcal H}}
\def\calI{{\mathcal I}}
\def\calJ{{\mathcal J}}
\def\calK{{\mathcal K}}
\def\calL{{\mathcal L}}
\def\calM{{\mathcal M}}
\def\calN{{\mathcal N}}
\def\calO{{\mathcal O}}
\def\calP{{\mathcal P}}
\def\calQ{{\mathcal Q}}
\def\calR{{\mathcal R}}
\def\calS{{\mathcal S}}
\def\calT{{\mathcal T}}
\def\calU{{\mathcal U}}
\def\calV{{\mathcal V}}
\def\calW{{\mathcal W}}
\def\calX{{\mathcal X}}
\def\calY{{\mathcal Y}}
\def\calZ{{\mathcal Z}}

\def\bA{\mbox {\boldmath $A$}}
\def\bB{\mbox {\boldmath $B$}}
\def\bC{\mbox {\boldmath $C$}}
\def\bD{\mbox {\boldmath $D$}}
\def\bE{\mbox {\boldmath $E$}}
\def\bF{\mbox {\boldmath $F$}}
\def\bG{\mbox {\boldmath $G$}}
\def\bH{\mbox {\boldmath $H$}}
\def\bI{\mbox {\boldmath $I$}}
\def\bJ{\mbox {\boldmath $J$}}
\def\bK{\mbox {\boldmath $K$}}
\def\bL{\mbox {\boldmath $L$}}
\def\bM{\mbox {\boldmath $M$}}
\def\bN{\mbox {\boldmath $N$}}
\def\bO{\mbox {\boldmath $O$}}
\def\bP{\mbox {\boldmath $P$}}
\def\bQ{\mbox {\boldmath $Q$}}
\def\bR{\mbox {\boldmath $R$}}
\def\bS{\mbox {\boldmath $S$}}
\def\bT{\mbox {\boldmath $T$}}
\def\bU{\mbox {\boldmath $U$}}
\def\bV{\mbox {\boldmath $V$}}
\def\bW{\mbox {\boldmath $W$}}
\def\bX{\mbox {\boldmath $X$}}
\def\bY{\mbox {\boldmath $Y$}}
\def\bZ{\mbox {\boldmath $Z$}}

\def\ba{\mbox {$\bf{a}$}}
\def\bb{\mbox {\boldmath $b$}}
\def\bc{\mbox {\boldmath $c$}}
\def\bd{\mbox {\boldmath $d$}}
\def\be{\mbox {\boldmath $e$}}
\def\bg{\mbox {\boldmath $g$}}
\def\bh{\mbox {\boldmath $h$}}
\def\bi{\mbox {\boldmath $i$}}
\def\bj{\mbox {\boldmath $j$}}
\def\bk{\mbox {\boldmath $k$}}
\def\bl{\mbox {\boldmath $l$}}
\def\bm{\mbox {\boldmath $m$}}
\def\bn{\mbox {\boldmath $n$}}
\def\bo{\mbox {\boldmath $o$}}
\def\bp{\mbox {\boldmath $p$}}
\def\bq{\mbox {\boldmath $q$}}
\def\br{\mbox {\boldmath $r$}}
\def\bs{\mbox {\boldmath $s$}}
\def\bt{\mbox {\boldmath $t$}}
\def\bu{\mbox {\boldmath $u$}}
\def\bv{\mbox {\boldmath $v$}}
\def\bw{\mbox {\boldmath $w$}}
\def\bx{\mbox {\boldmath $x$}}
\def\by{\mbox {\boldmath $y$}}
\def\bz{\mbox {\boldmath $z$}}

\newcommand{\snr}{\textup{SNR}}
\newcommand{\UE}{\mathrm{UE}}
\newcommand{\BS}{\mathrm{BS}}
\newcommand{\Passoc}{p_{_{I^{(1)}}}}
\newcommand{\Pintra}{p_{_{I^{(2)}}}}
\newcommand{\Pinter}{p_{_{I^{(3)}}}}

\newenvironment{Ex}
{\begin{adjustwidth}{0.04\linewidth}{0cm}
\begingroup\small
\vspace{-1.0em}
\raisebox{-.2em}{\rule{\linewidth}{0.3pt}}
\begin{example}
}
{
\end{example}
\vspace{-5mm}
\rule{\linewidth}{0.3pt}
\endgroup
\end{adjustwidth}}


\makeatletter
\let\save@mathaccent\mathaccent
\newcommand*\if@single[3]{%
  \setbox0\hbox{${\mathaccent"0362{#1}}^H$}%
  \setbox2\hbox{${\mathaccent"0362{\kern0pt#1}}^H$}%
  \ifdim\ht0=\ht2 #3\else #2\fi
  }
\newcommand*\rel@kern[1]{\kern#1\dimexpr\macc@kerna}
\newcommand*\widebar[1]{\@ifnextchar^{{\wide@bar{#1}{0}}}{\wide@bar{#1}{1}}}
\newcommand*\wide@bar[2]{\if@single{#1}{\wide@bar@{#1}{#2}{1}}{\wide@bar@{#1}{#2}{2}}}
\newcommand*\wide@bar@[3]{%
  \begingroup
  \def\mathaccent##1##2{%
    \let\mathaccent\save@mathaccent
    \if#32 \let\macc@nucleus\first@char \fi
    \setbox\z@\hbox{$\macc@style{\macc@nucleus}_{}$}%
    \setbox\tw@\hbox{$\macc@style{\macc@nucleus}{}_{}$}%
    \dimen@\wd\tw@
    \advance\dimen@-\wd\z@
    \divide\dimen@ 3
    \@tempdima\wd\tw@
    \advance\@tempdima-\scriptspace
    \divide\@tempdima 10
    \advance\dimen@-\@tempdima
    \ifdim\dimen@>\z@ \dimen@0pt\fi
    \rel@kern{0.6}\kern-\dimen@
    \if#31
      \overline{\rel@kern{-0.6}\kern\dimen@\macc@nucleus\rel@kern{0.4}\kern\dimen@}%
      \advance\dimen@0.4\dimexpr\macc@kerna
      \let\final@kern#2%
      \ifdim\dimen@<\z@ \let\final@kern1\fi
      \if\final@kern1 \kern-\dimen@\fi
    \else
      \overline{\rel@kern{-0.6}\kern\dimen@#1}%
    \fi
  }%
  \macc@depth\@ne
  \let\math@bgroup\@empty \let\math@egroup\macc@set@skewchar
  \mathsurround\z@ \frozen@everymath{\mathgroup\macc@group\relax}%
  \macc@set@skewchar\relax
  \let\mathaccentV\macc@nested@a
  \if#31
    \macc@nested@a\relax111{#1}%
  \else
    \def\gobble@till@marker##1\endmarker{}%
    \futurelet\first@char\gobble@till@marker#1\endmarker
    \ifcat\noexpand\first@char A\else
      \def\first@char{}%
    \fi
    \macc@nested@a\relax111{\first@char}%
  \fi
  \endgroup
}
\makeatother

\def\herm{\mathsf{H}}
\def\trans{\mathsf{T}}
\newcommand{\call}[1]{{\textsf{\small \textsc{#1}}}}
\newcommand{\callf}[1]{{\textsf{\footnotesize \textsc{#1}}}}

\def\argmax{\mathrm{arg}\max}
\def\argmin{\mathrm{arg}\min}
\renewcommand{\algorithmicrequire}{\textbf{Input:}}
\renewcommand{\algorithmicensure}{\textbf{Output:}}
\algdef{SE}[PROCEDURE]{Procedure}{EndProcedure}%
   [2]{\algorithmicprocedure\ \textproc{#1}\ifthenelse{\equal{#2}{}}{}{(#2)}}%
   {\algorithmicend\ \algorithmicprocedure}%
\algdef{SE}[FUNCTION]{Function}{EndFunction}%
   [2]{\algorithmicfunction\ \textproc{#1}\ifthenelse{\equal{#2}{}}{}{(#2)}}%
   {\algorithmicend\ \algorithmicfunction}%


\newcommand{\Hossein}[1]{{\textcolor{blue}{\emph{**Hossein: #1**}}}}
\newcommand{\Gabor}[1]{{\textcolor{cyan}{\emph{**Gabor: #1**}}}}
\newcommand{\Hadi}[1]{{\textcolor{red}{#1}}}
\newcommand{\gf}[1]{{\textcolor{cyan}{#1}}}
\newcommand{\Rev}[1]{{\textcolor{blue}{#1}}}
\newcommand{\Extended}[1]{{\color{green!50!black}{#1}}}


\begin{acronym}
  \acro{2G}{Second Generation}
  \acro{3G}{3$^\text{rd}$~Generation}
  \acro{3GPP}{3$^\text{rd}$~Generation Partnership Project}
  \acro{4G}{4$^\text{th}$~Generation}
  \acro{5G}{5$^\text{th}$~Generation}
  \acro{AA}{Antenna Array}
  \acro{AC}{Admission Control}
  \acro{AD}{Attack-Decay}
  \acro{ADSL}{Asymmetric Digital Subscriber Line}
	\acro{AHW}{Alternate Hop-and-Wait}
  \acro{AMC}{Adaptive Modulation and Coding}
	\acro{AP}{Access Point}
  \acro{APA}{Adaptive Power Allocation}
  \acro{ARMA}{Autoregressive Moving Average}
  \acro{ATES}{Adaptive Throughput-based Efficiency-Satisfaction Trade-Off}
  \acro{AWGN}{additive white Gaussian noise}
  \acro{BB}{Branch and Bound}
  \acro{BCD}{block-coordinate descent}
  \acro{BD}{block-diagonalization}
  \acro{BER}{Bit Error Rate}
  \acro{BF}{Best Fit}
  \acro{BLER}{BLock Error Rate}
  \acro{BPC}{Binary power control}
  \acro{BPSK}{Binary Phase-Shift Keying}
  \acro{BPA}{Best \ac{PDPR} Algorithm}
  \acro{BRA}{Balanced Random Allocation}
  \acro{BS}{base station}
  \acro{CAP}{Combinatorial Allocation Problem}
  \acro{CAPEX}{Capital Expenditure}
  \acro{CBF}{Coordinated Beamforming}
  \acro{CBR}{Constant Bit Rate}
  \acro{CBS}{Class Based Scheduling}
  \acro{CC}{Congestion Control}
  \acro{CDF}{Cumulative Distribution Function}
  \acro{CDMA}{Code-Division Multiple Access}
  \acro{CI}{coherence interval}
  \acro{CL}{Closed Loop}
  \acro{CLPC}{Closed Loop Power Control}
  \acro{CNR}{Channel-to-Noise Ratio}
  \acro{CPA}{Cellular Protection Algorithm}
  \acro{CPICH}{Common Pilot Channel}
  \acro{CoMP}{Coordinated Multi-Point}
  \acro{CQI}{Channel Quality Indicator}
  \acro{CRM}{Constrained Rate Maximization}
	\acro{CRN}{Cognitive Radio Network}
  \acro{CS}{Coordinated Scheduling}
  \acro{CSI}{channel state information}
  \acro{CSIR}{channel state information at the receiver}
  \acro{CSIT}{channel state information at the transmitter}
  \acro{CUE}{cellular user equipment}
  \acro{D2D}{device-to-device}
  \acro{DCA}{Dynamic Channel Allocation}
  \acro{DE}{Differential Evolution}
  \acro{DFT}{Discrete Fourier Transform}
  \acro{DIST}{Distance}
  \acro{DL}{downlink}
  \acro{DMA}{Double Moving Average}
	\acro{DMRS}{Demodulation Reference Signal}
  \acro{D2DM}{D2D Mode}
  \acro{DMS}{D2D Mode Selection}
  \acro{DPC}{Dirty Paper Coding}
  \acro{DRA}{Dynamic Resource Assignment}
  \acro{DSA}{Dynamic Spectrum Access}
  \acro{DSM}{Delay-based Satisfaction Maximization}
  \acro{ECC}{Electronic Communications Committee}
  \acro{EFLC}{Error Feedback Based Load Control}
  \acro{EI}{Efficiency Indicator}
  \acro{eNB}{Evolved Node B}
  \acro{EPA}{Equal Power Allocation}
  \acro{EPC}{Evolved Packet Core}
  \acro{EPS}{Evolved Packet System}
  \acro{E-UTRAN}{Evolved Universal Terrestrial Radio Access Network}
  \acro{ES}{Exhaustive Search}
  \acro{FDD}{frequency division duplexing}
  \acro{FDM}{Frequency Division Multiplexing}
  \acro{FER}{Frame Erasure Rate}
  \acro{FF}{Fast Fading}
  \acro{FSB}{Fixed Switched Beamforming}
  \acro{FST}{Fixed SNR Target}
  \acro{FTP}{File Transfer Protocol}
  \acro{GA}{Genetic Algorithm}
  \acro{GBR}{Guaranteed Bit Rate}
  \acro{GLR}{Gain to Leakage Ratio}
  \acro{GOS}{Generated Orthogonal Sequence}
  \acro{GPL}{GNU General Public License}
  \acro{GRP}{Grouping}
  \acro{HARQ}{Hybrid Automatic Repeat Request}
  \acro{HMS}{Harmonic Mode Selection}
  \acro{HOL}{Head Of Line}
  \acro{HSDPA}{High-Speed Downlink Packet Access}
  \acro{HSPA}{High Speed Packet Access}
  \acro{HTTP}{HyperText Transfer Protocol}
  \acro{ICMP}{Internet Control Message Protocol}
  \acro{ICI}{Intercell Interference}
  \acro{ID}{Identification}
  \acro{IETF}{Internet Engineering Task Force}
  \acro{ILP}{Integer Linear Program}
  \acro{JRAPAP}{Joint RB Assignment and Power Allocation Problem}
  \acro{UID}{Unique Identification}
  \acro{IID}{Independent and Identically Distributed}
  \acro{IIR}{Infinite Impulse Response}
  \acro{ILP}{Integer Linear Problem}
  \acro{IMT}{International Mobile Telecommunications}
  \acro{INV}{Inverted Norm-based Grouping}
	\acro{IoT}{Internet of Things}
  \acro{IP}{Internet Protocol}
  \acro{IPv6}{Internet Protocol Version 6}
  \acro{ISD}{Inter-Site Distance}
  \acro{ISI}{Inter Symbol Interference}
  \acro{ITU}{International Telecommunication Union}
  \acro{JOAS}{Joint Opportunistic Assignment and Scheduling}
  \acro{JOS}{Joint Opportunistic Scheduling}
  \acro{JP}{Joint Processing}
	\acro{JS}{Jump-Stay}
  \acro{KKT}{Karush-Kuhn-Tucker}
  \acro{L3}{Layer-3}
  \acro{LAC}{Link Admission Control}
  \acro{LA}{Link Adaptation}
  \acro{LC}{Load Control}
  \acro{LOS}{Line of Sight}
  \acro{LP}{Linear Programming}
  \acro{LS}{least squares}
  \acro{LTE}{Long Term Evolution}
  \acro{LTE-A}{LTE-Advanced}
  \acro{LTE-Advanced}{Long Term Evolution Advanced}
  \acro{M2M}{Machine-to-Machine}
  \acro{MAB}{multi-armed bandit}
  \acro{MAC}{medium access control}
  \acro{MANET}{Mobile Ad hoc Network}
  \acro{MC}{Modular Clock}
  \acro{MCS}{Modulation and Coding Scheme}
  \acro{MDB}{Measured Delay Based}
  \acro{MDI}{Minimum D2D Interference}
  \acro{MF}{Matched Filter}
  \acro{MG}{Maximum Gain}
  \acro{MH}{Multi-Hop}
  \acro{MIMO}{multiple input multiple output}
  \acro{MINLP}{Mixed Integer Nonlinear Programming}
  \acro{MIP}{Mixed Integer Programming}
  \acro{MISO}{Multiple Input Single Output}
  \acro{MLWDF}{Modified Largest Weighted Delay First}
  \acro{MME}{Mobility Management Entity}
  \acro{MMSE}{minimum mean squared error}
  \acro{MOS}{Mean Opinion Score}
  \acro{MPF}{Multicarrier Proportional Fair}
  \acro{MRA}{Maximum Rate Allocation}
  \acro{MR}{Maximum Rate}
  \acro{MRC}{Maximum Ratio Combining}
  \acro{MRT}{Maximum Ratio Transmission}
  \acro{MRUS}{Maximum Rate with User Satisfaction}
  \acro{MS}{mobile station}
  \acro{MSE}{mean squared error}
  \acro{MSI}{Multi-Stream Interference}
  \acro{MTC}{Machine-Type Communication}
  \acro{MTSI}{Multimedia Telephony Services over IMS}
  \acro{MTSM}{Modified Throughput-based Satisfaction Maximization}
  \acro{MU-MIMO}{multiuser multiple input multiple output}
  \acro{MU}{multi-user}
  \acro{NAS}{Non-Access Stratum}
  \acro{NB}{Node B}
  \acro{NE}{Nash equilibrium}
	\acro{NCL}{Neighbor Cell List}
  \acro{NLP}{Nonlinear Programming}
  \acro{NLOS}{Non-Line of Sight}
  \acro{NMSE}{Normalized Mean Square Error}
  \acro{NORM}{Normalized Projection-based Grouping}
  \acro{NP}{Non-Polynomial Time}
  \acro{NRT}{Non-Real Time}
  \acro{NSPS}{National Security and Public Safety Services}
  \acro{O2I}{Outdoor to Indoor}
  \acro{OFDMA}{orthogonal frequency division multiple access}
  \acro{OFDM}{orthogonal frequency division multiplexing}
  \acro{OFPC}{Open Loop with Fractional Path Loss Compensation}
	\acro{O2I}{Outdoor-to-Indoor}
  \acro{OL}{Open Loop}
  \acro{OLPC}{Open-Loop Power Control}
  \acro{OL-PC}{Open-Loop Power Control}
  \acro{OPEX}{Operational Expenditure}
  \acro{ORB}{Orthogonal Random Beamforming}
  \acro{JO-PF}{Joint Opportunistic Proportional Fair}
  \acro{OSI}{Open Systems Interconnection}
  \acro{PAIR}{D2D Pair Gain-based Grouping}
  \acro{PAPR}{Peak-to-Average Power Ratio}
  \acro{P2P}{Peer-to-Peer}
  \acro{PC}{Power Control}
  \acro{PCI}{Physical Cell ID}
  \acro{PDF}{Probability Density Function}
  \acro{PDPR}{pilot-to-data power ratio}
  \acro{PER}{Packet Error Rate}
  \acro{PF}{Proportional Fair}
  \acro{P-GW}{Packet Data Network Gateway}
  \acro{PL}{Pathloss}
  \acro{PPR}{pilot power ratio}
  \acro{PRB}{Physical Resource Block}
  \acro{PROJ}{Projection-based Grouping}
  \acro{ProSe}{Proximity Services}
  \acro{PS}{Packet Scheduling}
  \acro{PSO}{Particle Swarm Optimization}
  \acro{PZF}{Projected Zero-Forcing}
  \acro{QAM}{Quadrature Amplitude Modulation}
  \acro{QoS}{Quality of Service}
  \acro{QPSK}{Quadri-Phase Shift Keying}
  \acro{RAISES}{Reallocation-based Assignment for Improved Spectral Efficiency and Satisfaction}
  \acro{RAN}{Radio Access Network}
  \acro{RA}{Resource Allocation}
  \acro{RAT}{Radio Access Technology}
  \acro{RATE}{Rate-based}
  \acro{RB}{resource block}
  \acro{RBG}{Resource Block Group}
  \acro{REF}{Reference Grouping}
  \acro{RLC}{Radio Link Control}
  \acro{RM}{Rate Maximization}
  \acro{RNC}{Radio Network Controller}
  \acro{RND}{Random Grouping}
  \acro{RRA}{Radio Resource Allocation}
  \acro{RRM}{Radio Resource Management}
  \acro{RSCP}{Received Signal Code Power}
  \acro{RSRP}{Reference Signal Receive Power}
  \acro{RSRQ}{Reference Signal Receive Quality}
  \acro{RR}{Round Robin}
  \acro{RRC}{Radio Resource Control}
  \acro{RSSI}{Received Signal Strength Indicator}
  \acro{RT}{Real Time}
  \acro{RU}{Resource Unit}
  \acro{RUNE}{RUdimentary Network Emulator}
  \acro{RV}{Random Variable}
  \acro{RZF}{regularized zero forcing}
  \acro{SAC}{Session Admission Control}
  \acro{SCM}{Spatial Channel Model}
  \acro{SC-FDMA}{Single Carrier - Frequency Division Multiple Access}
  \acro{SD}{Soft Dropping}
  \acro{S-D}{Source-Destination}
  \acro{SDPC}{Soft Dropping Power Control}
  \acro{SDMA}{Space-Division Multiple Access}
  \acro{SER}{Symbol Error Rate}
  \acro{SES}{Simple Exponential Smoothing}
  \acro{S-GW}{Serving Gateway}
  \acro{SINR}{signal-to-interference-plus-noise ratio}
  \acro{SI}{Satisfaction Indicator}
  \acro{SIP}{Session Initiation Protocol}
  \acro{SISO}{single input single output}
  \acro{SIMO}{Single Input Multiple Output}
  \acro{SIR}{signal-to-interference ratio}
  \acro{SLNR}{Signal-to-Leakage-plus-Noise Ratio}
  \acro{SMA}{Simple Moving Average}
  \acro{SNR}{signal-to-noise ratio}
  \acro{SORA}{Satisfaction Oriented Resource Allocation}
  \acro{SORA-NRT}{Satisfaction-Oriented Resource Allocation for Non-Real Time Services}
  \acro{SORA-RT}{Satisfaction-Oriented Resource Allocation for Real Time Services}
  \acro{SPF}{Single-Carrier Proportional Fair}
  \acro{SRA}{Sequential Removal Algorithm}
  \acro{SRS}{Sounding Reference Signal}
  \acro{SU-MIMO}{Single-User Multiple Input Multiple Output}
  \acro{SU}{Single-User}
  \acro{SVD}{Singular Value Decomposition}
  \acro{TCP}{Transmission Control Protocol}
  \acro{TDD}{time division duplexing}
  \acro{TDMA}{Time Division Multiple Access}
  \acro{TETRA}{Terrestrial Trunked Radio}
  \acro{TP}{Transmit Power}
  \acro{TPC}{Transmit Power Control}
  \acro{TTI}{Transmission Time Interval}
  \acro{TTR}{Time-To-Rendezvous}
  \acro{TSM}{Throughput-based Satisfaction Maximization}
  \acro{TU}{Typical Urban}
  \acro{UE}{user equipment}
  \acro{UEPS}{Urgency and Efficiency-based Packet Scheduling}
  \acro{UL}{uplink}
  \acro{ULA}{uniform linear array}
  \acro{UMTS}{Universal Mobile Telecommunications System}
  \acro{URI}{Uniform Resource Identifier}
  \acro{URM}{Unconstrained Rate Maximization}
  \acro{UT}{user terminal}
  \acro{VR}{Virtual Resource}
  \acro{VoIP}{Voice over IP}
  \acro{WAN}{Wireless Access Network}
  \acro{WCDMA}{Wideband Code Division Multiple Access}
  \acro{WF}{Water-filling}
  \acro{WiMAX}{Worldwide Interoperability for Microwave Access}
  \acro{WINNER}{Wireless World Initiative New Radio}
  \acro{WLAN}{Wireless Local Area Network}
  \acro{WMPF}{Weighted Multicarrier Proportional Fair}
  \acro{WPF}{Weighted Proportional Fair}
  \acro{WSN}{Wireless Sensor Network}
  \acro{WWW}{World Wide Web}
  \acro{XIXO}{(Single or Multiple) Input (Single or Multiple) Output}
  \acro{ZF}{zero-forcing}
  \acro{ZMCSCG}{Zero Mean Circularly Symmetric Complex Gaussian}
\end{acronym}

\maketitle

\vspace{-10mm}

\begin{abstract}
Inter-operator spectrum sharing in millimeter-wave bands has the potential of substantially increasing the spectrum utilization and providing a larger bandwidth to individual user equipment at the expense of increasing inter-operator interference.
Unfortunately, traditional model-based spectrum sharing schemes make idealistic assumptions about inter-operator coordination mechanisms in terms of latency and protocol overhead, while being sensitive to missing channel state information. In this paper, we propose hybrid model-based and data-driven multi-operator spectrum sharing mechanisms, which incorporate model-based beamforming and user association complemented by data-driven model refinements. Our solution has the same computational complexity as a model-based approach but has the major advantage of having substantially less signaling overhead. We discuss how limited channel state information and quantized codebook-based beamforming affect the learning and the spectrum sharing performance. We show that the proposed hybrid sharing scheme significantly improves spectrum utilization under realistic assumptions on inter-operator coordination and channel state information acquisition.
\end{abstract}


\section{Introduction}\label{sec: Introduction}
Millimeter-wave (mmWave) communications appear as a promising solution to support extremely high data rates
and low latency services in future wireless networks~\cite{Jiang2018LowLatency}.
Although mmWave bands offer a much wider spectrum than the commonly used sub~6-GHz bands,
it is still essential to seek an optimal use of the spectrum with the ultimate goal of maximizing the benefits
for users while fostering healthy competition in the spectrum market \cite{FCC:2016}.
Spectrum sharing addresses these goals by allowing multiple service providers (hereafter called operators)
to access the same band for the same or different uses.
This paper investigates the case of spectrum sharing for mobile broadband services among multiple mobile operators~\cite{Boccardi:16, Rebato:17, Hu:18, Jurdi:18, Doyle:14}.

Spectrum sharing provides substantially more bandwidth to individual operators but gives rise to increased interference levels.
This is usually addressed by heavy coordination among the \acp{BS} and computationally-prohibitive optimization problems.
In mmWave networks, however, large antenna arrays, directional communications, and the unique propagation environment
substantially simplify the problem of managing interference in a shared spectrum, making it more feasible~\cite{shokri2016Spectrum}.

\subsection{Literature Survey}
A series of recent works proposed various technology enablers and performance evaluation methods
that help realize the vision of managing the spectrum without bounds and networks without borders \cite{Doyle:14},
and ultimately making the best use of radio spectrum, see references \cite{Tsiftsis:16,Hu:18}
and references therein. In particular, Hu \textit{et al.} \cite{Hu:18} conducted a comprehensive survey on the benefits of spectrum sharing in four application scenarios of future wireless networks: wider coverage, massive capacity, massive connectivity, and low latency.

Rebato~\textit{et al.}~\cite{Rebato:17} proposed a hybrid spectrum sharing scheme in mmWave networks,
where an operator has exclusive access to some parts of the mmWave bands but also some shared access to some other mmWave bands.
The authors showed the advantages of this hybrid method (where data packets are scheduled through two mmWave carriers with different propagation characteristics) over traditional fully licensed or fully pooled spectrum access schemes.
Jurdi \textit{et al.}~\cite{Jurdi:18} used a system-level analysis to show that infrastructure sharing can be advantageously
combined with sharing spectrum licenses in the mmWave bands.

Coordination mechanisms have a large impact on the gains that spectrum sharing can achieve, and are intertwined
with the supporting architectural solutions~\cite{Hu:18,Mihovska:09, McMenamy:14, Holland:15, shokri2016Spectrum, Krysz:16, Boccardi:16}.
The early work by Mihovska \textit{et al.} proposed an approach for both intra- and inter-operator coordination scenarios and concluded that operators can advantageously pool spectrum resources when network loads are temporarily uneven among the cooperating operators \cite{Mihovska:09}.
Ghadikolaei \textit{et al.}~\cite{shokri2016Spectrum} showed that the large antenna setting can reduce the need for inter-operator coordination.
In fact, they showed that in the case of digital beamforming with ideal channel estimations
inter-operator coordination can be limited to cell-edge users.

A large part of the literature utilizes the increasing number of antennas to form narrow beams,
which reduces both intra- and inter-operator interference, defined as the interference within the
same or among different operators.
However, the inherent imperfections in terms of errors in the \ac{CSI} acquisition,
hardware limitations, and the constraints of quantized code-books make inter-operator coordination
a necessary ingredient of managing a common spectrum pool \cite{Xiao:16, shokri2016Spectrum}.
Besides, there is no consensus on how to properly model the coordination cost.

Due to the complexity and inherent data acquisition difficulties of coordinating a
large set of radio network nodes, learning-based coordination mechanisms to better manage the spectrum sharing
were recently proposed
by \cite{Hossain:14, Zhang:15, Srinivasan:16}.
Unfortunately, the schemes developed in \cite{Hossain:14} and \cite{Zhang:15} suit secondary users
and are not directly applicable in inter-operator spectrum sharing scenarios,
in which the participating operators share the spectrum pool on an equal right basis.
In contrast, the Q-learning framework of \cite{Srinivasan:16} facilitates inter-operator
sharing by the mechanism of intelligent user offloading.
However, none of these schemes addresses the problem of optimizing the network utility while maintaining an acceptable level of coordination and setting the precoders and combiners to reduce the intra- and inter-operator interference.

\subsection{Model-based Approaches for Spectrum Sharing}
Model-based approaches, while being ubiquitous in communication systems~\cite{tse2005fundamentals},
may rely on inaccurate and unrealistic assumptions for the sake of mathematical tractability.
Consequently, performance evaluation and protocol development based on such approximated and inaccurate models
run the risk of not working well in practice~\cite{Shokri2018IMSindex}.
Data-driven approaches address this disadvantage by learning and optimizing from the data
-- usually acquired by measurements --
making minimal assumptions on the system model.
These approaches have been the core of the success of modern machine learning and artificial intelligence.
Data-driven approaches, however, may need a large number of training samples to perform well,
which are hard to obtain in most wireless networks due to their inherent non-stationary nature~\cite{Sevakula:15}. This is indeed the case for general network optimization problems and in particular for spectrum sharing~\cite{gai2012combinatorial}.

In this paper, we advocate the use of a hybrid approach for spectrum sharing, in which the model-based part operates on a small timescale, whilst the data-driven part operates on a coarser time scale and refines the models used in the model-based part. The benefit of hybrid approaches has been demonstrated in the context of speech signal processing for the localization and tracking tasks~\cite{laufer2018hybrid} and in these parallel and independent works \cite{zappone2018model,zappone2019wireless}.

\subsection{Contributions of the Present Paper}
In this paper, we propose a framework to analyze and quantify the benefits of spectrum sharing over exclusive spectrum access for a multi-operator millimeter-wave network. More specifically, we capture the trade-offs among the signaling cost, coordination complexity, and overall network performance by an optimization task that takes as input a model for the rate functions and returns the optimal association and coordination policies throughout the network along with proper beamforming vectors.
We then augment this approach by adding a learning functionality that continuously refines the rate models to compensate for missing information (mostly missing \ac{CSI}) and to keep the signaling overhead manageable.
To enable this new function, every operator runs some carefully designed rate measurement tasks, reports the results to a cloud server that keeps an updated dataset for the learning and runs the spectrum sharing optimization problem using the updated data-driven rate models.
The main contributions of our work can be summarized as follows:
\begin{itemize}
\item
We propose a new generic and tractable approach for modeling the cost of coordination among multiple \acp{BS},
which is of significant interest on itself, beyond the scope of this paper.

\item 
We investigate the gains of beamforming and coordination for spectrum sharing schemes in mmWave networks.
We argue that a pure model-based solution approach to this problem is infeasible,
mainly due to modeling inaccuracy, the overhead of pilot transmission, and the lack of sufficient information
(including erroneous or completely missing \ac{CSI}).

\item
We develop a hybrid model-based and data-driven approach where the model-based part optimizes
the decision variables (association and coordination) and finds proper beamforming vectors, 
and the data-driven part sequentially and continuously refines the model. Our approach has the same computational complexity as the pure model-based approach but operates with a much lower signaling overhead.\footnote{Among other differences,``hybrid'' in~\cite{Rebato:17} refers to the scheduling of the data packets through two different carriers whereas our ``hybrid'' refers to the joint use of model-based and data-driven approached for spectrum sharing at the mmWave bands, leading to completely different design principles.}


\item We then use domain-specific knowledge (large antenna arrays and the sparse scattering environment of mmWave systems) to properly initialize the learning process to minimize its running complexity while guaranteeing the user performance.

\item We discuss how large antenna arrays, limited feedback, and imperfect/missing \ac{CSI} affect the learning process and consequently the spectrum sharing performance.
\end{itemize}
Conceptually, our hybrid solution could be considered both in a centralized and in a more realistic distributed implementation.

\subsection{Paper Organization}
The rest of the paper is organized as follows. We introduce our system model, including a novel coordination model,
in Section~\ref{sec: system-model}. We formulate the problem of spectrum sharing in Section~\ref{sec: model-based-solution} and discuss the complexities of pure model-based approaches.
Section~\ref{sec: hybrid-driven-solution} develops our hybrid solution approach and numerical performance evaluations.
We provide important engineering insights in Section~\ref{sec: discussions}, followed by concluding remarks of Section~\ref{sec: concluding-remarks}. Due to space limitations, we have provided all the proofs and extended numerical results in the extended version of this paper~\cite{shokri2019learhningSPECSreport}.

\textit{Notations:} Capital bold letters denote matrices and lower bold letters denote vectors. The superscripts $(\bX)^\trans$, $(\bX)^{\herm}$, $(\bX)^\dag$ stand for the transpose, transpose conjugate, and Moore-Penrose pseudo-inverse of $\bX$, respectively. The subscript $[\bX]_{mn}$ denotes entry of $\bX$ at row $m$ and column $n$, and $[\bX]_{n}$ represents column $n$ of $\bX$. $\bI_x$, and $\mathbf{1}_x$, and $\mathbf{0}_{x}$ are the identity, all-one, and all-zero matrices of size $x$, respectively. Table~\ref{table: notations} lists the main symbols used in the paper.

\section{System Model}
\label{sec: system-model}
In this paper, we use the following system model for our model-based approach that we propose in Section~\ref{sec: model-based-solution}.
This system model is generic and embraces distinct model elements for the network, the employed association scheme,
the deployed antenna and channel models, and models for beamforming and multi-operator coordination.

\subsection{Network Model}
We consider the downlink of a multi-operator cellular network with a total bandwidth $W$
to be shared among $Z$ operators in the network.
Each operator $z$ controls and operates the subset $\calB_z$ of the BSs
such that $\calB = \calB_1 \cup \calB_1 \cup \ldots  \cup \calB_{Z}$ is the set of all BSs in the network.
With no infrastructure sharing, for example, $\{\calB_z\}_{z=1}^{Z}$ are disjoint sets.
We denote by $\calU$ the set of all UEs, by $\calU_z$ the set of all UEs of operator $z$, and by $W_z$ the bandwidth of operator $z$.
Without loss of generality, we assume universal frequency reuse within an operator's network.
Consequently,  all non-serving BSs of an operator cause interference to every \ac{UE} of that operator in the downlink.
\begin{table}[t]
  \centering
  \caption{Summary of main notations.}\label{table: notations}

\renewcommand{\arraystretch}{1.1}
  {
   \begin{tabular}{|*{8}{l|}}
\hline
   \textbf{Symbol} & \textbf{Definition} \\ \hline
    $b,i$ & Indices denoting a BS \\
    $u,j$ & Indices denoting a UE \\
    $k,z$ & Indices denoting an operator \\ \hline
    $N_{\BS},N_{\UE}$ & Number of antennas at every \ac{BS} and \ac{UE} \\
    $N_{b}$ & Number of UEs that are associated to BS $b$ \\
    $N_{b  u}$ & Number of paths between BS $b$ and UE $u$ \\
    $Z$ & Number of operators \\ \hline
    $\calU, \calB$ & Set of all UEs and BSs of all operators \\
    $\calU_{z},\calB_{z}$ & Set of UEs and BSs of operator $z$ \\
    $\mathcal{A}_{b}$ & Set of UEs that are associated to BS $b$ \\ \hline
    $W_{z}$ & Bandwidth of operator $z$ \\ \hline
    $\bA, \bC$ & Association and coordination matrices\\
    $\bP$ & Penalty matrix\\
    $p_b$ & Coordination penalty of BS $b$ for its UEs \\
\hline
    $L_{b  u}$ & Path loss between BS $b$ and UE $u$ \\
    \multirow{2}{*}{$\bH_{b  u}$} & \multirow{2}{*}{Channel matrix between BS $b$ and UE $u$} \\[1.5mm]
    & including large and small scale fading \\
    $\overline{\bH}_{b}$ & Effective channel from the perspective of BS $b$ \\
    $\ba_{\UE} (\theta)~$ & Antenna response of UEs to $\theta$ \\
    $\ba_{\BS} (\theta)~$ & Antenna response of BSs to $\theta$ \\\hline
    $\bw_{b u}^{\BS} $ & Precoding vector of BS $b$ when serving UE $u$\\
    $\bw_{u}^{\UE} $ & Combiner vector of UE $u$ \\ \hline
    $r_{u}$ & Long-term rate of UE $u$ \\ \hline
    $\rho^{\mathrm{Rx}}_{bu}$ & Received power of UE $u$ from BS $b$\\
    $I_{bu}^{(1)}$ & Received interference of UE $u$ from BS $b$ \\
    $I_{bu}^{(2)}$ & Intra-operator interference at UE $u$\\
    $I_{bu}^{(3)}$ & Inter-operator interference at UE $u$\\ \hline
\end{tabular}}
\end{table}

\subsection{Association Model}\label{sec: association-model}
We denote by $a_{b  u}$ a binary variable that is equal to 1 if \ac{UE} $u \in \calU$ is served by  (or associated to) BS $b \in \calB$. We collect all binary control variables $a_{b u}$ in association matrix $\bA$, where $\bA = \sum_{z \in [Z]}\bA_z$. Binary matrix $\bA_z$ of size $|\calB| \times |\calU|$ is the association of operator $z$, namely $[\bA]_{bu} = 1$ if and only if $u\in\calU_z$ and $a_{bu}=1$.
Let $N_{b} = \sum_{u \in \calU}{a_{b u}}$ and $\mathcal{A}_{b}$ be the number and the set of UEs that are being served by BS $b$, respectively.
We also call $N_{b}$ the load of BS $b$.
Note that without national roaming, each BS can serve only UEs of the same operator.
Namely, $a_{b u} = 0$ for all $b \in \calB_z, u \in \calU_k$ where $z\neq k$.
We first impose the constraint that national roaming is not permitted, which will be relaxed in Section~\ref{sec: numerical-results-model-based1} to examine the potential performance improvement due to national roaming.


We define the association period as a consecutive series of \acp{CI} over which association $\bA$ remains unchanged,
see Fig.~\ref{fig: SuperframeStructure}.
Although beamforming should be recomputed every \ac{CI}, the association is a long-term process in the sense that it remains fixed over some \acp{CI}~\cite{andrews2014overview}. Such an assumption is natural, due to the inherent cost of handover for re-association.
In this paper, we investigate the performance of optimal association; i.e., we find the optimal $\bA_z$ for all operators.
Using these associations, the BSs and UEs recalculate their beamforming vectors every \ac{CI}. To avoid the interplay between the short-term scheduling and the association problem, which should be handled at different time scales, we ensure that each BS can serve all its associated UEs simultaneously by imposing that the number of served UEs is compatible with the number of RF chains at each BS.

\begin{figure}[t]
  \centering
  \includegraphics[width=0.8\columnwidth]{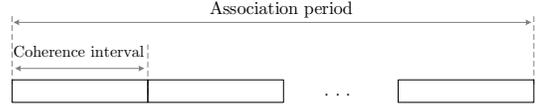}

  \caption{A UE-BS association period. Beamforming vectors are fixed only for one \ac{CI}, and should be recomputed afterward. The UE-BS association is fixed over a block of many \ac{CI} intervals, denoted as association period.
  }\label{fig: SuperframeStructure}
\end{figure}

\subsection{Antenna and Channel Model}\label{sec: ChannelModel}
We consider a half wavelength \ac{ULA}
of $N_{\BS}$ antenna elements for all BSs and a \ac{ULA} of $N_{\UE}$ antennas for all UEs,
albeit our mathematical framework can be easily extended to other antenna models.
We consider a narrowband mmWave channel model~\cite{Akdeniz2014MillimeterWave}.
Let $N_{b u}$ be the number of paths between BS $b \in \calB$ and UE $u  \in \calU$,
and $g_{b u n}$ be the complex gain of the $n$-th path that includes both path loss and small scale fading.
In particular, $g_{b u n}$ is a zero-mean complex Gaussian random variable with $\mathbb{E}[|g_{b u n}|^2] = L_{b u}$ for $n=1,2,\ldots,N_{b u}$,
where $L_{b u}$ is the path loss between BS $b$ and UE $u$.
The channel matrix between BS $b$ and UE $u$ is given by
\begin{equation}\label{eq: channel-matrix}
\bH_{b u} = \sqrt{\frac{N_{\BS}N_{\UE}}{N_{b u}}}\sum\limits_{n=1}^{N_{b u}}{g_{b u n} \, \ba_{\UE}\left(\theta_{b u n}^{\UE}\right) \ba_{\BS}^{\herm} \left(\theta_{b u n}^{\BS}\right)} \:,
\end{equation}
where $\ba_{\BS} \in \mathds{C}^{N_{\BS}}$ and $\ba_{\UE} \in \mathds{C}^{N_{\UE}}$ are the vector response functions of the BSs' and UEs' antenna arrays to the angles of arrival and departure (AoAs and AoDs), $\theta_{b u n}^{\BS}$ is the AoD of the $n$-th path,
$\theta_{b u n}^{\UE}$ is the AoA of the $n$-th path, and $(\cdot)^{\herm}$ is the conjugate transpose operator.
For a ULA with half wavelength antenna spacing at the BS, we have
\begin{equation}\label{eq: ULA-antenna-response}
\ba_{\BS}\left(\theta\right) = \frac{1}{\sqrt{N_{\BS}}}\left[1, e^{j \pi \sin(\theta)}, \ldots, e^{j (N_{\BS}-1)\pi\sin(\theta)}\right]^{\herm} \:.
\end{equation}
$\ba_{\UE}\left(\theta\right)$ can be obtained from~\eqref{eq: ULA-antenna-response} by changing $N_{\BS}$ to $N_{\UE}$.


\subsection{Beamforming and Coordination Models}\label{sec:Beamforming-and-coordination-model}
\subsubsection{Analog Combiners}
To simplify the implementation requirements, we consider an analog combiner using phase shifters at the UE side (only one RF chain per UE). With these phase shifters, each UE can only change its antenna boresight.
Let $\bw_{u}^{\UE} \in \mathbb{C}^{N_{\UE}}$ be the combining vector of UE $u$. Assume that BS $b$ serves UE $u$ and that the estimates of the channel gains and the corresponding AoAs are available. We pick for UE $u$ the analog combiner that maximizes its link budget~\cite{Ayach2012Capacity}, namely

\begin{equation}\label{eq: AnalogCombiner}
\bw_{u}^{\UE} = \ba_{\UE}\left(\theta_{b u n^{\star}}^{\UE}\right), ~~ \mbox{where} ~~ n^{\star} = \arg\max_n |g_{b u n}| \:.
\end{equation}

\subsubsection{Precoders}
For the sake of presentation simplicity, we assume that each BS employs a fully-digital precoder. At the end of this subsection, we show how to extend our derivations to the case of hybrid (analog-digital) precoding.

We assume that all the UEs are concurrently served by their respective BSs with multiuser MIMO. To ensure this, we impose the condition ${N_{b} \leq N_{\BS}}$ for all BSs and all operators in the next sections.
Let $\bW_{b}^{\BS} \in \mathbb{C}^{N_{\BS}\times N_{b}}$ be the digital precoding matrix at BS $b$ whose $u$-th column $\bw_{b u}^{\BS} \in \mathds{C}^{N_{\BS}}$ is the precoding vector for UE $u$.
We define the transmitted symbols of BS $b$ by $\sqrt{\lambda_{b}} \bW_{b}^{\BS}\bd_{b}$, where $\bd_{b} \in \mathbb{C}^{N_{b}}$ are the data symbols for the $N_{b}$ UEs of this cell with normalized power, and $\rho^{\mathrm{Tx}}$ is the average transmit power at each BS.
Moreover, $\lambda_{b}$ normalizes the maximum transmit power of the BS $b$ to $\rho$, namely
\begin{equation}\label{eq: power-normalization-precoder}
\lambda_{b} = \rho^{\mathrm{Tx}}/\mathrm{tr} \Big(\bW_{b}^{\BS} \left(\bW_{b}^{\BS}\right)^{\herm}\Big) \:.
\end{equation}

We consider \ac{RZF}, which is of practical {interest} for minimizing the inter-BS
(within and among different operators) interference.\footnote{We can replace RZF by almost any approach, e.g., \ac{MMSE}. Moreover, note that we do not require joint transmission, which may be infeasible if BSs belong to different operators, due to the latency involved in signaling through the corresponding core networks.}
For every BS $b$, define $\overline{\bH}_b$ as the effective channel that the digital precoder observes containing $\left(\bw_{u}^{\UE}\right)^{\herm}\bH_{b u}$ for several $u$ in its rows; formally defined later in this section.

Suppose that UE $u$ is being served by BS $b$, and that $\left(\bw_{u}^{\UE}\right)^{\herm}\bH_{b u}$ has appeared in row $m$ of $\overline{\bH}_b$.
Using \ac{RZF}, the precoding vector of UE $u$ is
\begin{equation}\label{eq: DigitalBeamforming}
\bw_{b u}^{\BS} = \left[\left(\overline{\bH}_b + \delta
\begin{bmatrix} \bI \\ \mathbf{0} \end{bmatrix}
\right)^{\dag}\right]_{m} \:,
\end{equation}
where $\delta$ is an arbitrary (usually very small) positive number, and $\bI$ is an identity matrix of proper size.

In the case of hybrid precoding, we can still design $\bW_b^{\BS}$ based on~\eqref{eq: DigitalBeamforming}
and then approximate the true hybrid precoding matrix with a cascade of an analog and a digital precoder, while satisfying the constant-modulus constraint of the analog precoder; see, e.g.,~\cite{Yu_HybridAM_16} and \cite{Ghauch_SED_16}.

\subsubsection{Coordination}
Let UE $u$ be served by BS $b$ using combiner $\bw_{u}^{\UE}$.
Define the effective channel between any BS $i \in \calB$ and any UE $u$ as $\overline{\bH}_{i u}:= \left(\bw_{u}^{\UE}\right)^{\herm}\bH_{iu} $. In fact, the effective channel is the actual channel between BS $i$ and UE $u$ processed by the analog combiner of the UE. We define binary matrix $\bC \in \{0,1\}^{|\calB| \times |\calU|}$ where $[\bC]_{iu} = 1$ if and only if BS $i\in\calB$ can estimate the effective channel $\overline{\bH}_{i u}$. If $i=b$, then acquiring this effective channel has a much lower cost than if $i$ and $b$ belong to different operators. To model this, we add a penalty for the coordination to promote the optimal use of coordinations. For a given association of the BSs and UEs $\bA$, we assign a penalty $[\bP]_{iu}$ corresponding to the element $[\bC]_{iu}$ of the coordination matrix. For sake of simplicity, in the following, we consider a constant penalty matrix $\bP$, though it can be in general a function of the distance, operator load, and number of antennas, among others. The penalty terms may vary for each operator, reflecting various billing policies.

When UE $u\in\calU_z$ is associated with BS $b\in\calB_z$,
we may have $0 \leq [\bP]_{bu} < [\bP]_{iu} < [\bP]_{ju}$,
where $i \in \calB_z\setminus\{b\}$ and $j\in \calB_k$, $k\neq z$,
incurring almost no cost of estimating the channel of the own served UEs,
a higher cost of estimating the effective channel of a UE within operator, and an even higher cost of estimating
the effective channel of UEs of other operators.
This abstraction of the penalty matrix facilitates the cross-layer design of spectrum sharing. Notice there should be some inter-operator architectural support whose design is out of the scope of this paper. Interested readers are referred to \cite{Boccardi:16} and references therein.
To implement the penalty matrix, we recall the set of BSs and UEs of all operators.
Furthermore, the penalty matrix $\bP_0$ represents the cost associated with channel estimation,
where $[\bP_0]_{bu}$ is the penalty when BS $b$ estimates the channel of UE $u$.
This penalty may not be identical for all non-serving operator.
\begin{remark}
Let $\mathbf{1}_{M\times N}$ be an all-one matrix of size $M\times N$, and $\mathrm{blkdiag}(\cdot)$ denote a mapping of the arguments to a block diagonal matrix. By setting $\bC = \bA$, $\bC = \mathrm{blkdiag}(\mathbf{1}_{|\calB_1|\times |\calU_1|},\ldots,\mathbf{1}_{|\calB_Z|\times |\calU_Z|})$, and $\bC=\mathbf{1}_{|\calB|\times |\calU|}$, our approach can model ``no coordination,'' ``partial coordination,'' and ``full coordination'' scenarios of~\cite{shokri2016Spectrum}, respectively.
\end{remark}
\begin{example} Let $p$ and $\widebar{p}$ denote the penalty of intra-operator and inter-operator coordinations,
incurring identical costs for all operators.
The template penalty is then computed as
\begin{equation}\label{eq: P0}
[\bP_0]_{bu} = \begin{cases}
                     p, & \mbox{if~} b\in\calB_z \mbox{~and~}u\in\calU_z \\
                     \bar{p}, & \mbox{if~} b\in\calB\setminus\calB_z \mbox{~and~}u\in\calU_z.
                   \end{cases}
\end{equation}
\end{example}

Given $\bP_0$, we then set $[\bP]_{bu} \leftarrow [\bP_0]_{bu} + a_{bu}\left(p_b - [\bP_0]_{bu}\right)$ for any $b\in \calB$, where $a_{bu} \in \{0,1\}$, $u \in \calU$, and $p_b$ is the coordination penalty for the UEs associated to BS $b$. Note that $a_{bu}=1$ means BS $b$ serves UE $u$.
The coordination cost in each \ac{CI} to serve users of operator $z$ is thus
\begin{equation}\label{eq: coordination-cost}
\sum_{u\in\calU_z}\sum_{b\in\calB}~[\bC]_{bu} [\bP]_{bu} \:.
\end{equation}
Ultimately, the goal is to find the optimal coordination policy
that maximizes a network objective (e.g., sum-rate of the \acp{UE})
while bounding the coordination cost;
see Section~\ref{sec: model-based-solution}.
As we show throughout this paper, under realistic settings for \ac{CSI} acquisitions and network topologies, this optimization task is possible by a hybrid model-based and data-driven approach.


The effective channel $\overline{\bH}_i$ is a matrix of dimension $ \sum_{u}[\bC]_{iu}\times N_{\BS}$
whose rows correspond to the effective channels $\overline{\bH}_{i u}$ for $\{u\mid [\bC]_{iu}=1\}$.

\begin{example}
To illustrate the notations of this paper, Fig.~\ref{fig: ToyExample} shows an illustrative example with two operators,
each having 2 BSs and 5 UEs. We run this example throughout the paper.
Every BS can estimate the effective channel of its associated UEs.
BS 2 can estimate the effective channel toward UE 5 via intra-operator coordination.
Moreover, BSs 1 and 2 (of operator blue) can estimate their effective channel toward UE 7 (of operator red).
For this topology, ${\calB_1 = \{1,2\}}, {\calB_2 = \{3,4\}}, N_b = 2 ~\mbox{(for all $b$)}, {\calU_1 = \{1,\ldots,5\}}$, and ${\calU_2 = \{6,\ldots,10\}}$.
For all $b\in \calB$ and $u\in \calU$, the penalty of coordinating with associated UEs is $p_b = 1$,
while the penalty of intra-operator and inter-operator coordination is 10 and 100, respectively.

\begin{figure}[t]
  \centering
 {\footnotesize\begin{tikzpicture}
\begin{axis}[%
width=0.6\columnwidth,
height=0.45\columnwidth,
at={(0\columnwidth,0\columnwidth)},
scale only axis,
xmin=0,
xmax=270,
ymin=0,
ymax=200,
axis background/.style={fill=white},
ymajorgrids
]
\addplot [color=black, line width=2.3pt]
  table[row sep=crcr]{%
100	90\\
20	10\\
};

\addplot [color=black, line width=2.3pt]
  table[row sep=crcr]{%
100	90\\
45	135\\
};

\addplot [color=black, line width=2.3pt]
  table[row sep=crcr]{%
175	90\\
165	140\\
};

\addplot [color=black, line width=2.3pt]
  table[row sep=crcr]{%
175	90\\
200	20\\
};

\addplot [color=black, line width=2.3pt]
  table[row sep=crcr]{%
100	90\\
120	55\\
};

\addplot [color=black, line width=2.3pt]
  table[row sep=crcr]{%
140	125\\
98	119\\
};

\addplot [color=black, line width=2.3pt]
  table[row sep=crcr]{%
140	125\\
135	85\\
};

\addplot [color=black, line width=2.3pt]
  table[row sep=crcr]{%
200	150\\
220	107\\
};

\addplot [color=black, line width=2.3pt]
  table[row sep=crcr]{%
200	150\\
185	185\\
};

\addplot [color=black, line width=2.3pt]
  table[row sep=crcr]{%
200	150\\
230	125\\
};

\addplot [color=green, line width=1.0pt]
  table[row sep=crcr]{%
100	90\\
20	10\\
};

\addplot [color=green, line width=1.0pt]
  table[row sep=crcr]{%
100	90\\
45	135\\
};

\addplot [color=green, line width=1.0pt]
  table[row sep=crcr]{%
100	90\\
120	55\\
};

\addplot [color=green, line width=1.0pt]
  table[row sep=crcr]{%
100	90\\
135	85\\
};

\addplot [color=green, line width=1.0pt]
  table[row sep=crcr]{%
175	90\\
165	140\\
};

\addplot [color=green, line width=1.0pt]
  table[row sep=crcr]{%
175	90\\
200	20\\
};

\addplot [color=green, line width=1.0pt]
  table[row sep=crcr]{%
175	90\\
120	55\\
};

\addplot [color=green, line width=1.0pt]
  table[row sep=crcr]{%
175	90\\
135	85\\
};

\addplot [color=green, line width=1.0pt]
  table[row sep=crcr]{%
140	125\\
98	119\\
};

\addplot [color=green, line width=1.0pt]
  table[row sep=crcr]{%
140	125\\
135	85\\
};

\addplot [color=green, line width=1.0pt]
  table[row sep=crcr]{%
200	150\\
220	107\\
};

\addplot [color=green, line width=1.0pt]
  table[row sep=crcr]{%
200	150\\
185	185\\
};

\addplot [color=green, line width=1.0pt]
  table[row sep=crcr]{%
200	150\\
230	125\\
};

\node[font=\bfseries,draw=blue,rectangle,fill=blue,minimum size=4mm,font=\color{white}] at (100,90){\textbf{1}};
\node[font=\bfseries,draw=blue,rectangle,fill=blue,minimum size=4mm,font=\color{white}] at (175,90){\textbf{2}};
\node[font=\bfseries,draw=red,rectangle,fill=red,minimum size=4mm,font=\color{white}] at (140,125){\textbf{3}};
\node[font=\bfseries,draw=red,rectangle,fill=red,minimum size=4mm,font=\color{white}] at (200,150){\textbf{4}};

\addplot [color=blue, draw=none, mark size=5.0pt, mark=*, mark options={solid, fill=blue, blue}]
  table[row sep=crcr]{%
20	10\\
45	135\\
165	140\\
200	20\\
120	55\\
};

\addplot [color=red, draw=none, mark size=5.0pt, mark=*, mark options={solid, fill=red, red}]
  table[row sep=crcr]{%
98	119\\
135	85\\
220	107\\
185	185\\
230	125\\
};

\node[right, align=left]
at (axis cs:26.25,10) {1};
\node[right, align=left]
at (axis cs:51.25,135) {2};
\node[right, align=left]
at (axis cs:171.25,140) {3};
\node[right, align=left]
at (axis cs:206.25,20) {4};
\node[right, align=left]
at (axis cs:126.25,55) {5};
\node[right, align=left]
at (axis cs:99,130) {6};
\node[right, align=left]
at (axis cs:141.25,91) {7};
\node[right, align=left]
at (axis cs:226.25,107) {8};
\node[right, align=left]
at (axis cs:191.25,185) {9};
\node[right, align=left]
at (axis cs:236.25,125) {10};

\end{axis}
\end{tikzpicture}%
}

  \caption{An example topology with two operators, red and blue. BSs and UEs are marked by squares and circles, respectively. Black lines show association. Green lines show coordination.}
  \label{fig: ToyExample}
\end{figure}
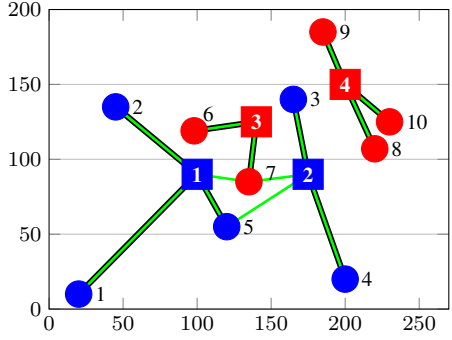
\begin{align*}
& \bA  = \begin{pmatrix}
          1 & 1 & 0 & 0 & 1 & 0 & 0 & 0 & 0 & 0 \\
          0 & 0 & 1 & 1 & 0 & 0 & 0 & 0 & 0 & 0 \\
          0 & 0 & 0 & 0 & 0 & 1 & 1 & 0 & 0 & 0 \\
          0 & 0 & 0 & 0 & 0 & 0 & 0 & 1 & 1 & 1
        \end{pmatrix}\:, \\
& \bC = \begin{pmatrix}
          1 & 1 & 0 & 0 & 1 & 0 & 1 & 0 & 0 & 0 \\
          0 & 0 & 1 & 1 & 1 & 0 & 1 & 0 & 0 & 0 \\
          0 & 0 & 0 & 0 & 0 & 1 & 1 & 0 & 0 & 0 \\
          0 & 0 & 0 & 0 & 0 & 0 & 0 & 1 & 1 & 1
        \end{pmatrix} \:, \\
& \bP =
{\setlength{\arraycolsep}{4pt}
\begin{pmatrix}
          1 & 1 & 0 & 0 & 1 & 0 & 100 & 0 & 0 & 0 \\
          0 & 0 & 1 & 1 & 10 & 0 & 100 & 0 & 0 & 0 \\
          0 & 0 & 0 & 0 & 0 & 1 & 1 & 0 & 0 & 0 \\
          0 & 0 & 0 & 0 & 0 & 0 & 0 & 1 & 1 & 1
        \end{pmatrix}
}\:, \\
\overline{\bH}_{1} & = \begin{pmatrix}
                       \left(\bw_{1}^{\UE}\right)^{\herm}\bH_{11} \\
                       \left(\bw_{2}^{\UE}\right)^{\herm}\bH_{12} \\
                       \left(\bw_{5}^{\UE}\right)^{\herm}\bH_{15} \\
                       \left(\bw_{7}^{\UE}\right)^{\herm}\bH_{17}
                     \end{pmatrix} \:,
\overline{\bH}_{2} = \begin{pmatrix}
                       \left(\bw_{3}^{\UE}\right)^{\herm}\bH_{23} \\
                       \left(\bw_{4}^{\UE}\right)^{\herm}\bH_{23} \\
                       \left(\bw_{5}^{\UE}\right)^{\herm}\bH_{25} \\
                       \left(\bw_{7}^{\UE}\right)^{\herm}\bH_{27}
                     \end{pmatrix} \:, 
\end{align*}
\begin{align*}
\overline{\bH}_{3} & = \begin{pmatrix}
                       \left(\bw_{6}^{\UE}\right)^{\herm}\bH_{36} \\
                       \left(\bw_{7}^{\UE}\right)^{\herm}\bH_{37}
                     \end{pmatrix} \:,
\overline{\bH}_{4} = \begin{pmatrix}
                       \left(\bw_{8}^{\UE}\right)^{\herm}\bH_{48} \\
                       \left(\bw_{9}^{\UE}\right)^{\herm}\bH_{49} \\
                       \left(\bw_{10}^{\UE}\right)^{\herm}\bH_{4,10} \\
                     \end{pmatrix}\:,
\end{align*}
leading to a total coordination penalty of 215 for operator 1 and 5 for operator 2.
\end{example}
Given $\bA$, we can find $\bw_{u}^{\UE}$ from \eqref{eq: AnalogCombiner}.
Then, given a coordination matrix $\bC$, every BS $i$ obtains $\overline{\bH}_i$ and finds $\bw_{b u}^{\BS}$ from \eqref{eq: DigitalBeamforming}.
The data transmission phase then follows.

\section{Problem Formulation and Solution Approaches}
\label{sec: model-based-solution}
In this section, we formulate the problem of spectrum sharing among multiple operators. Specifically, we use the models of Section~\ref{sec: system-model} and then show the complexity and limitations of this model-based approach to optimize the beamforming, association, and coordination for spectrum sharing.
Note that all the variables with superscript $b$ are operator-dependent.
This dependency exists since BS $b$ belongs to $\calB_z$ for some $z$.

\subsection{SINR and Rate for Model-based Approach}
We define a cell as the set of UEs that are served by the same BS.
The received power at each UE $u \in \calU_z$ when the serving BS is $b \in \calB$ consists of the desired power $\rho^{\mathrm{Rx}}$, intra-cell interference $I^{(1)}$, inter-cell interference $I^{(2)}$, inter-operator interference $I_{bu}^{(3)}$,
and noise power spectral density $\sigma^2$.
$I^{(1)}$ corresponds to the signals transmitted to other UEs by the same BS.
$I^{(2)}$ denotes the interference from the signals transmitted by other BSs of the same network operator.
$I_{bu}^{(3)}$ consists of the interference from the signals transmitted by all BSs of other operators $\calB\setminus \calB_z$ toward their own UEs.

We first note that the received power at UE $u$ from BS $b$ is
\begin{equation}\label{eq: Rx-power}
\rho^{\mathrm{Rx}}_{bu} = \lambda_b |\left(\bw_{u}^{\UE}\right)^{\herm} \bH_{b u} \bw_{b u}^{\BS}|^2 \:.
 \end{equation}
Recall the definitions of the binary association variables $a_{ij}$ and the set of associated UEs $\mathcal{A}_{i}$.
Each BS serves multiple UEs at the same time and frequency resources, as UEs are separable at the spatial domain.
The intra-cell and inter-cell interference to UE $u \in \calU_z$ when served by BS $b \in \calB_z$ are
\begin{equation}\label{eq: intra-cell-interference-digital-precoder}
I_{bu}^{(1)} = \lambda_{b}\sum\limits_{j\in\mathcal{A}_{b}\setminus\{u\}}\left| \left(\bw_{u}^{\UE}\right)^{\herm} \bH_{b u} \bw_{bj}^{\BS} \right|^2 \:,
\end{equation}
\begin{equation}\label{eq: inter-cell-interference-digital-precoder}
I_{bu}^{(2)} = \sum\limits_{i \in \calB_z \setminus \{b\}} \lambda_{i}\sum\limits_{j\in\mathcal{A}_{i}}\left| \left(\bw_{u}^{\UE}\right)^{\herm} \bH_{iu} \bw_{ij}^{\BS} \right|^2 \:.
\end{equation}

For UE $u$, inter-operator interference $I_{bu}^{(3)}$ depends on the set of operators (and BSs) that share the same bandwidth. 
Without loss of generality, we assume that $W_z = W$. With universal frequency reuse, UE $u$ receives interference from all BSs of all operators, and the inter-operator interference can be expressed as
\begin{equation}\label{eq: inter-operator-interference-digital-precoder}
I_{bu}^{(3)} = \sum\limits_{k=1 \hfill \atop k \neq z \hfill}^{Z} \, \sum\limits_{i \in \calB_k \setminus \{b\}} \lambda_{i}\sum\limits_{j\in\mathcal{A}_{i}}\left| \left(\bw_{u}^{\UE}\right)^{\herm} \bH_{iu} \bw_{i j}^{\BS} \right|^2 \:.
\end{equation}
Note that the special characteristics of mmWave networks, such as high penetration loss and directional communications, substantially reduce the interference components~\eqref{eq: intra-cell-interference-digital-precoder}--\eqref{eq: inter-operator-interference-digital-precoder}, compared to sub-6~GHz systems, as established in~\cite{di2014stochastic}. We use this property later on in Section~\ref{sec: hybrid-driven-solution} to substantially reduce the complexity of the hybrid model-based and data-driven optimization algorithm by a proper initialization.

The long-term rate that UE $u$ will receive from all BSs is
\begin{equation}\label{eq: rate}
r_u = \sum_{b \in \calB}{a_{b u}W_z \mathbb{E} \left[\log\left(1 + \frac{\rho^{\mathrm{Rx}}_{bu}}{I_{bu}^{(1)} + I_{bu}^{(2)} + I_{bu}^{(3)} + W_z\sigma^2}\right) \right]} \:,
\end{equation}
where the expectation is over all random channel gains. Notice that we do not assume joint transmission, so $\sum_{b\in \calB}a_{bu}=1$ for all $u\in\calU$.
Sharing the spectrum increases the bandwidth available to each operator (with a prelog contribution to the rate in high SINR regimes); however, it also increases the interference power.
As we discuss later in this section, not being able to compute $r_{u}$ due to missing \ac{CSI} is an important disadvantage of the model-based approaches.

\subsection{Optimal Spectrum Sharing with Model-based Approach}\label{sec: optimal-sharing-strategy}

For given $\bA$ and $\bC$, and in every \ac{CI},
BS $b$ estimates $\overline{\bH}_b$, and finds the digital precoding and analog combiner using~\eqref{eq: DigitalBeamforming}.
Given that each BS can evaluate the average rate $r_{u}$ for its associated UEs from~\eqref{eq: rate},
a cloud server (logical controller) collects $\{r_{u}\}$ from all BSs,
computes the coordination cost per \ac{CI} from~\eqref{eq: coordination-cost}, and evaluates a network utility $f_z(\bA,\bC)$ for operator $z$.
Given $r_u$ in \eqref{eq: rate}, we use a logarithmic utility that ensures both high network throughput and some level of fairness among individual UEs~\cite{andrews2014overview}:
\begin{equation}\label{eq: ObjectiveFunction1}
f_z = \sum\nolimits_{u \in \calU_z}\,{\log r_{u}} \:.
\end{equation}

Given $\calB$ and $\calU$, the controller computes $\bP_0$ from~\eqref{eq: P0}
and formulates the following optimization problem to find the optimal association and coordination strategies:
\begin{subequations}\label{eq: optimal-spectrum-sharing}
\begin{alignat}{3}
\label{eq: objective-P1}
\calP_1\hspace{-0.8mm}:~\hspace{-1.3mm} & \underset{\bA, \bC}{\mathrm{maximize}} \hspace{1.3mm} && \sum\nolimits_{z=1}^{Z}\alpha_z f_z (\bA,\bC) \:,
\\
\label{eq: const-unique-association}
& {\text{subject to}} &&
\sum\nolimits_{b \in \calB_z} a_{b u} = 1 \:, \hspace{1.5mm} \forall u \in \calU_z, 1\leq z \leq Z\:,
\\
\label{eq: cell-size-p1}
& && \sum\nolimits_{u \in \calU_z} a_{b u} \leq N_{\BS} \:, \hspace{0.1mm} \forall b \in \calB_z, 1\leq z \leq Z \:, \\
& &&
[\bP]_{bu} = [\bP_0]_{bu} + a_{bu}\left(p_b - [\bP_0]_{bu}\right) \\
\label{eq: coordination-budget}
& &&
\sum\nolimits_{u\in\calU_z}\sum\nolimits_{b\in\calB} [\bC]_{bu} [\bP]_{bu} \leq P_{z}^{\max}, \nonumber \\[-1mm]
& && \hspace{45mm} \forall 1 \leq z \leq Z \:, \\
\label{eq: binary-variables-11}
& &&
a_{b u}=0 \:, \forall b \in \calB_k, u \in \calU_z, k\neq z, 1 \leq z,k \leq Z, \\
\label{eq: binary-variables}
& &&
a_{b u} \in \{0,1 \} \:, c_{b u} \in \{0,1 \} \:, \forall b \in \calB, u \in \calU \:,
\end{alignat}
\end{subequations}
where $\{\alpha_z\}_z$ are a set of positive constants that scalarize the multi-objective optimization problem, and $\sum_{z=1}^{Z}\alpha_z = 1$. Constraint~\eqref{eq: const-unique-association} guarantees association of each UE to only one BS, mitigating joint scheduling requirements among BSs. Constraint~\eqref{eq: cell-size-p1} ensures that $N_{b} \leq N_{\BS}$,
so all $N_{b}$ UEs that are associated to BS $b$ can be served together with multiuser MIMO. If $N_{b} < N_{\BS} $,
some RF chains will be switched off, and the BS automatically gives higher transmit power to the active RF chains.
Constraint \eqref{eq: coordination-budget} ensures that the coordination cost of every operator is upper-bounded by its maximum budget $P_{z}^{\max}$.
Constraint~\eqref{eq: binary-variables-11} ensures that the UEs of operator $z$ can be only served by BSs of the same operator.

\begin{remark}[Signaling Complexity]\label{remark: sigcomplex}
To compute the rate function, \eqref{eq: objective-P1}, and thereby solving \eqref{eq: optimal-spectrum-sharing}, an operator should coordinate with all UEs, leading to a coordination cost of $\sum\nolimits_{u\in\calU}\sum\nolimits_{b\in\calB} [\bP]_{bu}$. Notice that ~\eqref{eq: coordination-budget} is the coordination cost of network operation when the solution to \eqref{eq: optimal-spectrum-sharing} is deployed.
\end{remark}


\emph{Special Case (National roaming variant of $\calP_1$):} We can modify $\calP_1$ to allow for national roaming. To this end, we should only replace \eqref{eq: const-unique-association} by $\sum_{b \in \calB} a_{b u} = 1 \:, \forall u \in \calU$, replace \eqref{eq: cell-size-p1} by  $\sum_{u \in \calU} a_{b u} \leq N_{\BS}, \forall b \in \calB$, and remove constraint~\eqref{eq: binary-variables-11}.

\emph{Special Case 2 (Distributed implementation of $\calP_1$):}
To allow for a distributed implementation of $\calP_1$, we enforce the following design constraints. First, each operator maximizes only its own utility $f_z$. Second, we do not allow for inter-operator coordination, namely $[\bC]_{bu}=0$ when $b\in \calB_z$ and $u \in \calU\setminus \calU_z$. Third, every BS $b\in\calB_z$ locally approximates the rate functions $r_{u}$ by a quantity $\widehat{r}_{u}$ that takes as input only $\bA_z$ and $[\bC]_{bu}$ for all $b\in\calB_z, u\in\calU_z$. Consequently, $f_z$ can be calculated without any inter-operator coordination. Now, it is straightforward to formulate a variation of $\calP_1$, which can be independently solved by individual operators in parallel without any inter-operator coordination. While technically possible, we do not use this distributed implementation in the rest of the paper.

\subsection{Practical Considerations for Model-based Approach}\label{sec: practical_considerations}
While theoretically sound, optimally solving $\calP_1$ (and its distributed variant) with the signaling and time-limitations of the conventional radio access and core networks would be infeasible.
To solve $\calP_1$, for instance, the BSs of every operator should be able to send (or receive) pilot signals to all UEs of all operators and exchange a huge amount of information with a central controller, which should then solve $\calP_1$.
The complexity and cost of such level of channel estimation and coordination grow large with the number of BSs and UEs,
and are in general overwhelming for mmWave networks with dense BS deployment.
Moreover, if BSs or UEs belong to different network operators, a huge inter-operator signaling via the core networks is required for synchronization and for the calculation of $I_{bu}^{(3)}$.
Furthermore, channel aging may render the exchanged information outdated before it serves its purpose.
To tackle this problem, most of the works in the literature consider the noise-limited assumption and ignore the interference terms, see~\cite{Yu2016Distributed} and references therein, namely $I_{bu}^{(2)} = I_{bu}^{(3)} \approx 0$.
This is a rather limiting assumption, and it has been shown that a few links may observe strong interference~\cite{Park2009Analysis}.
Moreover, with the interference-free assumption, there is no gain of using a precoder to reduce the interference,
which would be an incorrect design decision.

These impairments have prohibited the application of optimal spectrum sharing in state-of-the-art wireless systems.
Nonetheless, the solution of $\calP_1$ gives a theoretical upper bound for the performance of spectrum sharing (a benchmark).
In the following, we take a data-driven approach as a completely different alternative to address the problem of spectrum sharing in mmWave networks.



\subsection{Illustrative Numerical Results}\label{sec: numerical-results-model-based1}
In this section, we numerically investigate the effect of the input/design parameters, namely, the number of antennas, network topology, association, and coordination levels. We use these insights to develop an efficient hybrid approach in the next section.

We consider an illustrative scenario of two operators, each having 2 BSs and 10 UEs with the topology of Fig.~\ref{fig: ToyExample}. We generate 100 random channels, find the beamforming vectors in every realization, and evaluate the interference terms. We consider two antenna settings: $( N_{\BS} = 8, N_{\UE} = 2)$ and $( N_{\BS} = 64, N_{\UE} = 16)$.
For all $b$ and $u$, we set $p_b = 1$, the intra-operator coordination penalty to $10$, and the inter-operator coordination penalty to $100$.
Fig.~\ref{fig: Example1_1} shows three example settings for the association and coordination matrices.
In the first scenario, Fig.~\ref{fig: Example1_1_1}, we assume no coordination among \acp{UE} and unintended \acp{BS}, namely $\bC =\bA$.
In the second scenario, Fig.~\ref{fig: Example1_2_1}, we set  $[\bC]_{bu} = [\bA]_{bu} $ and then allow BS 1 to estimate the effective channel toward UE 6 and cancel the resulting interference.
In Fig.~\ref{fig: Example1_3_1}, we assume full coordination, namely $[\bC]_{bu} = [\bA]_{bu} $.
This level of coordination may improve the rate performance at the expense of a very high coordination cost.
Moreover, for every antenna setting, we run $\calP_1$ and its national roaming variant, introduced in Special Case of Section~\ref{sec: optimal-sharing-strategy}. Fig.~\ref{fig: Example1_4_1} shows the optimal association and coordination for $( N_{\BS} = 8, N_{\UE} = 2)$ and $P_z^{\max} = 120$ (up to one inter-operator coordination) with national roaming. To find this solution, we first apply a continuous relaxation to the binary constraints of $\calP_1$ and then rounding to recover binary solutions. Furthermore, we assume that $[\bA]_{bu} = 1$ implies $[\bC]_{bu} = 1$ for every $b\in\calB$ and $u \in \calU$, which further reduces the feasibility space. This is a natural simplification of the optimization problem, as a serving BS will always estimate the channels of its serving UEs.
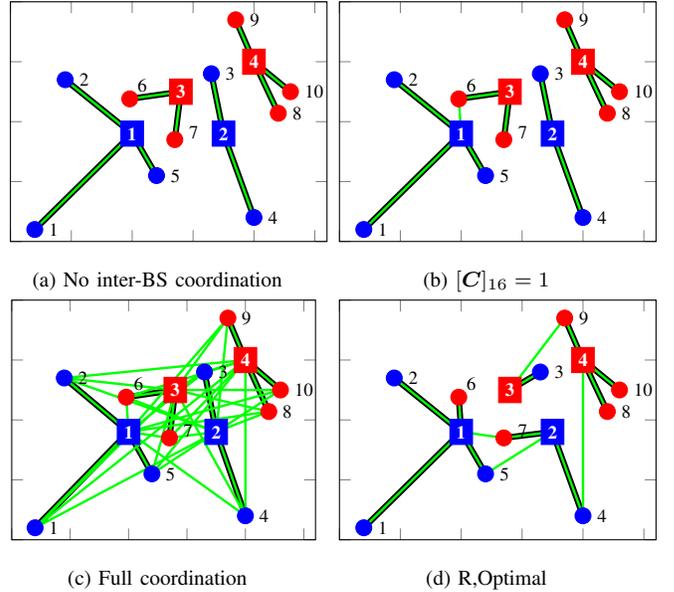
\begin{figure}[!t]
  \centering
\begin{minipage}{0.48\columnwidth}
  \centering
 {\scriptsize\begin{tikzpicture}
\begin{axis}[%
width=0.99\columnwidth,
height=0.75\columnwidth,
at={(0in,0in)},
scale only axis,
xmin=0,
xmax=260,
xtick={0,50,100,150,200,250},
xticklabels=\empty,
ymin=0,
ymax=200,
yticklabels=\empty,
axis background/.style={fill=white},
]
\addplot [color=black, line width=2.3pt]
  table[row sep=crcr]{%
100	90\\
20	10\\
};

\addplot [color=black, line width=2.3pt]
  table[row sep=crcr]{%
100	90\\
45	135\\
};

\addplot [color=black, line width=2.3pt]
  table[row sep=crcr]{%
175	90\\
165	140\\
};

\addplot [color=black, line width=2.3pt]
  table[row sep=crcr]{%
175	90\\
200	20\\
};

\addplot [color=black, line width=2.3pt]
  table[row sep=crcr]{%
100	90\\
120	55\\
};

\addplot [color=black, line width=2.3pt]
  table[row sep=crcr]{%
140	125\\
98	119\\
};

\addplot [color=black, line width=2.3pt]
  table[row sep=crcr]{%
140	125\\
135	85\\
};

\addplot [color=black, line width=2.3pt]
  table[row sep=crcr]{%
200	150\\
220	107\\
};

\addplot [color=black, line width=2.3pt]
  table[row sep=crcr]{%
200	150\\
185	185\\
};

\addplot [color=black, line width=2.3pt]
  table[row sep=crcr]{%
200	150\\
230	125\\
};

\addplot [color=green, line width=0.9pt]
  table[row sep=crcr]{%
100	90\\
20	10\\
};

\addplot [color=green, line width=0.9pt]
  table[row sep=crcr]{%
100	90\\
45	135\\
};

\addplot [color=green, line width=0.9pt]
  table[row sep=crcr]{%
100	90\\
120	55\\
};

\addplot [color=green, line width=0.9pt]
  table[row sep=crcr]{%
175	90\\
165	140\\
};

\addplot [color=green, line width=0.9pt]
  table[row sep=crcr]{%
175	90\\
200	20\\
};

\addplot [color=green, line width=0.9pt]
  table[row sep=crcr]{%
140	125\\
98	119\\
};

\addplot [color=green, line width=0.9pt]
  table[row sep=crcr]{%
140	125\\
135	85\\
};

\addplot [color=green, line width=0.9pt]
  table[row sep=crcr]{%
200	150\\
220	107\\
};

\addplot [color=green, line width=0.9pt]
  table[row sep=crcr]{%
200	150\\
185	185\\
};

\addplot [color=green, line width=0.9pt]
  table[row sep=crcr]{%
200	150\\
230	125\\
};

\node[font=\bfseries,draw=blue,rectangle,fill=blue,minimum size=3mm,font=\color{white}] at (100,90){\textbf{1}};
\node[font=\bfseries,draw=blue,rectangle,fill=blue,minimum size=3mm,font=\color{white}] at (175,90){\textbf{2}};
\node[font=\bfseries,draw=red,rectangle,fill=red,minimum size=3mm,font=\color{white}] at (140,125){\textbf{3}};
\node[font=\bfseries,draw=red,rectangle,fill=red,minimum size=3mm,font=\color{white}] at (200,150){\textbf{4}};

\addplot [color=blue, draw=none, mark size=3.0pt, mark=*, mark options={solid, fill=blue, blue}]
  table[row sep=crcr]{%
20	10\\
45	135\\
165	140\\
200	20\\
120	55\\
};

\addplot [color=red, draw=none, mark size=3.0pt, mark=*, mark options={solid, fill=red, red}]
  table[row sep=crcr]{%
98	119\\
135	85\\
220	107\\
185	185\\
230	125\\
};

\node[right, align=left]
at (axis cs:26.25,10) {1};
\node[right, align=left]
at (axis cs:51.25,135) {2};
\node[right, align=left]
at (axis cs:171.25,140) {3};
\node[right, align=left]
at (axis cs:206.25,20) {4};
\node[right, align=left]
at (axis cs:126.25,55) {5};
\node[right, align=left]
at (axis cs:99,130) {6};
\node[right, align=left]
at (axis cs:141.25,91) {7};
\node[right, align=left]
at (axis cs:226.25,107) {8};
\node[right, align=left]
at (axis cs:191.25,185) {9};
\node[right, align=left]
at (axis cs:236.25,125) {10};

\end{axis}
\end{tikzpicture}%
}

  \subcaption{No inter-BS coordination}
  \label{fig: Example1_1_1}
\end{minipage}
\begin{minipage}{0.48\columnwidth}
  \centering
 {\scriptsize\begin{tikzpicture}
\begin{axis}[%
width=0.99\columnwidth,
height=0.75\columnwidth,
at={(0in,0in)},
scale only axis,
xmin=0,
xmax=260,
xtick={0,50,100,150,200,250},
xticklabels=\empty,
ymin=0,
ymax=200,
yticklabels=\empty,
axis background/.style={fill=white},
]
\addplot [color=black, line width=2.3pt]
  table[row sep=crcr]{%
100	90\\
20	10\\
};

\addplot [color=black, line width=2.3pt]
  table[row sep=crcr]{%
100	90\\
45	135\\
};

\addplot [color=black, line width=2.3pt]
  table[row sep=crcr]{%
175	90\\
165	140\\
};

\addplot [color=black, line width=2.3pt]
  table[row sep=crcr]{%
175	90\\
200	20\\
};

\addplot [color=black, line width=2.3pt]
  table[row sep=crcr]{%
100	90\\
120	55\\
};

\addplot [color=black, line width=2.3pt]
  table[row sep=crcr]{%
140	125\\
98	119\\
};

\addplot [color=black, line width=2.3pt]
  table[row sep=crcr]{%
140	125\\
135	85\\
};

\addplot [color=black, line width=2.3pt]
  table[row sep=crcr]{%
200	150\\
220	107\\
};

\addplot [color=black, line width=2.3pt]
  table[row sep=crcr]{%
200	150\\
185	185\\
};

\addplot [color=black, line width=2.3pt]
  table[row sep=crcr]{%
200	150\\
230	125\\
};

\addplot [color=green, line width=0.9pt]
  table[row sep=crcr]{%
100	90\\
20	10\\
};

\addplot [color=green, line width=0.9pt]
  table[row sep=crcr]{%
100	90\\
45	135\\
};

\addplot [color=green, line width=0.9pt]
  table[row sep=crcr]{%
100	90\\
120	55\\
};

\addplot [color=green, line width=0.9pt]
  table[row sep=crcr]{%
175	90\\
165	140\\
};

\addplot [color=green, line width=0.9pt]
  table[row sep=crcr]{%
175	90\\
200	20\\
};

\addplot [color=green, line width=0.9pt]
  table[row sep=crcr]{%
98	119\\
100 90\\
};

\addplot [color=green, line width=0.9pt]
  table[row sep=crcr]{%
140	125\\
98	119\\
};

\addplot [color=green, line width=0.9pt]
  table[row sep=crcr]{%
140	125\\
135	85\\
};

\addplot [color=green, line width=0.9pt]
  table[row sep=crcr]{%
200	150\\
220	107\\
};

\addplot [color=green, line width=0.9pt]
  table[row sep=crcr]{%
200	150\\
185	185\\
};

\addplot [color=green, line width=0.9pt]
  table[row sep=crcr]{%
200	150\\
230	125\\
};

\node[font=\bfseries,draw=blue,rectangle,fill=blue,minimum size=3mm,font=\color{white}] at (100,90){\textbf{1}};
\node[font=\bfseries,draw=blue,rectangle,fill=blue,minimum size=3mm,font=\color{white}] at (175,90){\textbf{2}};
\node[font=\bfseries,draw=red,rectangle,fill=red,minimum size=3mm,font=\color{white}] at (140,125){\textbf{3}};
\node[font=\bfseries,draw=red,rectangle,fill=red,minimum size=3mm,font=\color{white}] at (200,150){\textbf{4}};

\addplot [color=blue, draw=none, mark size=3.0pt, mark=*, mark options={solid, fill=blue, blue}]
  table[row sep=crcr]{%
20	10\\
45	135\\
165	140\\
200	20\\
120	55\\
};

\addplot [color=red, draw=none, mark size=3.0pt, mark=*, mark options={solid, fill=red, red}]
  table[row sep=crcr]{%
98	119\\
135	85\\
220	107\\
185	185\\
230	125\\
};

\node[right, align=left]
at (axis cs:26.25,10) {1};
\node[right, align=left]
at (axis cs:51.25,135) {2};
\node[right, align=left]
at (axis cs:171.25,140) {3};
\node[right, align=left]
at (axis cs:206.25,20) {4};
\node[right, align=left]
at (axis cs:126.25,55) {5};
\node[right, align=left]
at (axis cs:99,130) {6};
\node[right, align=left]
at (axis cs:141.25,91) {7};
\node[right, align=left]
at (axis cs:226.25,107) {8};
\node[right, align=left]
at (axis cs:191.25,185) {9};
\node[right, align=left]
at (axis cs:236.25,125) {10};

\end{axis}
\end{tikzpicture}%
}

  \subcaption{$[\bC]_{16} = 1$}
  \label{fig: Example1_2_1}
\end{minipage}
\begin{minipage}{0.48\columnwidth}
  \centering
 {\scriptsize\begin{tikzpicture}
\begin{axis}[%
width=0.95\columnwidth,
height=0.75\columnwidth,
at={(0in,0in)},
scale only axis,
xmin=0,
xmax=260,
xtick={0,50,100,150,200,250},
xticklabels=\empty,
ymin=0,
ymax=200,
yticklabels=\empty,
axis background/.style={fill=white},
]
\addplot [color=black, line width=2.3pt]
  table[row sep=crcr]{%
100	90\\
20	10\\
};

\addplot [color=black, line width=2.3pt]
  table[row sep=crcr]{%
100	90\\
45	135\\
};

\addplot [color=black, line width=2.3pt]
  table[row sep=crcr]{%
175	90\\
165	140\\
};

\addplot [color=black, line width=2.3pt]
  table[row sep=crcr]{%
175	90\\
200	20\\
};

\addplot [color=black, line width=2.3pt]
  table[row sep=crcr]{%
100	90\\
120	55\\
};

\addplot [color=black, line width=2.3pt]
  table[row sep=crcr]{%
140	125\\
98	119\\
};

\addplot [color=black, line width=2.3pt]
  table[row sep=crcr]{%
140	125\\
135	85\\
};

\addplot [color=black, line width=2.3pt]
  table[row sep=crcr]{%
200	150\\
220	107\\
};

\addplot [color=black, line width=2.3pt]
  table[row sep=crcr]{%
200	150\\
185	185\\
};

\addplot [color=black, line width=2.3pt]
  table[row sep=crcr]{%
200	150\\
230	125\\
};

\addplot [color=green, line width=0.9pt]
  table[row sep=crcr]{%
98	119\\
100 90\\
};

\addplot [color=green, line width=0.9pt]
  table[row sep=crcr]{%
98	119\\
175 90\\
};

\addplot [color=green, line width=0.9pt]
  table[row sep=crcr]{%
98	119\\
140 125\\
};

\addplot [color=green, line width=0.9pt]
  table[row sep=crcr]{%
98	119\\
200 150\\
};

\addplot [color=green, line width=0.9pt]
  table[row sep=crcr]{%
135	85\\
100 90\\
};

\addplot [color=green, line width=0.9pt]
  table[row sep=crcr]{%
135	85\\
175 90\\
};

\addplot [color=green, line width=0.9pt]
  table[row sep=crcr]{%
135	85\\
140 125\\
};

\addplot [color=green, line width=0.9pt]
  table[row sep=crcr]{%
135	85\\
200 150\\
};

\addplot [color=green, line width=0.9pt]
  table[row sep=crcr]{%
220	107\\
100 90\\
};

\addplot [color=green, line width=0.9pt]
  table[row sep=crcr]{%
220	107\\
175 90\\
};

\addplot [color=green, line width=0.9pt]
  table[row sep=crcr]{%
220	107\\
140 125\\
};

\addplot [color=green, line width=0.9pt]
  table[row sep=crcr]{%
220	107\\
200 150\\
};

\addplot [color=green, line width=0.9pt]
  table[row sep=crcr]{%
185	185\\
100 90\\
};

\addplot [color=green, line width=0.9pt]
  table[row sep=crcr]{%
185	185\\
175 90\\
};

\addplot [color=green, line width=0.9pt]
  table[row sep=crcr]{%
185	185\\
140 125\\
};

\addplot [color=green, line width=0.9pt]
  table[row sep=crcr]{%
185	185\\
200 150\\
};

\addplot [color=green, line width=0.9pt]
  table[row sep=crcr]{%
230	125\\
100 90\\
};
\addplot [color=green, line width=0.9pt]
  table[row sep=crcr]{%
230	125\\
175 90\\
};
\addplot [color=green, line width=0.9pt]
  table[row sep=crcr]{%
230	125\\
140 125\\
};
\addplot [color=green, line width=0.9pt]
  table[row sep=crcr]{%
230	125\\
200 150\\
};

\addplot [color=green, line width=0.9pt]
  table[row sep=crcr]{%
20	10\\
100 90\\
};
\addplot [color=green, line width=0.9pt]
  table[row sep=crcr]{%
20	10\\
175 90\\
};
\addplot [color=green, line width=0.9pt]
  table[row sep=crcr]{%
20	10\\
140 125\\
};
\addplot [color=green, line width=0.9pt]
  table[row sep=crcr]{%
20	10\\
200 150\\
};

\addplot [color=green, line width=0.9pt]
  table[row sep=crcr]{%
45	135\\
100 90\\
};
\addplot [color=green, line width=0.9pt]
  table[row sep=crcr]{%
45	135\\
175 90\\
};
\addplot [color=green, line width=0.9pt]
  table[row sep=crcr]{%
45	135\\
140 125\\
};
\addplot [color=green, line width=0.9pt]
  table[row sep=crcr]{%
45	135\\
200 150\\
};

\addplot [color=green, line width=0.9pt]
  table[row sep=crcr]{%
165	140\\
100 90\\
};
\addplot [color=green, line width=0.9pt]
  table[row sep=crcr]{%
165	140\\
175 90\\
};
\addplot [color=green, line width=0.9pt]
  table[row sep=crcr]{%
165	140\\
140 125\\
};
\addplot [color=green, line width=0.9pt]
  table[row sep=crcr]{%
165	140\\
200 150\\
};

\addplot [color=green, line width=0.9pt]
  table[row sep=crcr]{%
200	20\\
100 90\\
};
\addplot [color=green, line width=0.9pt]
  table[row sep=crcr]{%
200	20\\
175 90\\
};
\addplot [color=green, line width=0.9pt]
  table[row sep=crcr]{%
200	20\\
140 125\\
};
\addplot [color=green, line width=0.9pt]
  table[row sep=crcr]{%
200	20\\
200 150\\
};

\addplot [color=green, line width=0.9pt]
  table[row sep=crcr]{%
120	55\\
100 90\\
};
\addplot [color=green, line width=0.9pt]
  table[row sep=crcr]{%
120	55\\
175 90\\
};
\addplot [color=green, line width=0.9pt]
  table[row sep=crcr]{%
120	55\\
140 125\\
};
\addplot [color=green, line width=0.9pt]
  table[row sep=crcr]{%
120	55\\
200 150\\
};

\node[font=\bfseries,draw=blue,rectangle,fill=blue,minimum size=3mm,font=\color{white}] at (100,90){\textbf{1}};
\node[font=\bfseries,draw=blue,rectangle,fill=blue,minimum size=3mm,font=\color{white}] at (175,90){\textbf{2}};
\node[font=\bfseries,draw=red,rectangle,fill=red,minimum size=3mm,font=\color{white}] at (140,125){\textbf{3}};
\node[font=\bfseries,draw=red,rectangle,fill=red,minimum size=3mm,font=\color{white}] at (200,150){\textbf{4}};

\addplot [color=blue, draw=none, mark size=3.0pt, mark=*, mark options={solid, fill=blue, blue}]
  table[row sep=crcr]{%
20	10\\
45	135\\
165	140\\
200	20\\
120	55\\
};

\addplot [color=red, draw=none, mark size=3.0pt, mark=*, mark options={solid, fill=red, red}]
  table[row sep=crcr]{%
98	119\\
135	85\\
220	107\\
185	185\\
230	125\\
};

\node[right, align=left]
at (axis cs:26.25,10) {1};
\node[right, align=left]
at (axis cs:51.25,135) {2};
\node[right, align=left]
at (axis cs:171.25,140) {3};
\node[right, align=left]
at (axis cs:206.25,20) {4};
\node[right, align=left]
at (axis cs:126.25,55) {5};
\node[right, align=left]
at (axis cs:99,130) {6};
\node[right, align=left]
at (axis cs:141.25,91) {7};
\node[right, align=left]
at (axis cs:226.25,107) {8};
\node[right, align=left]
at (axis cs:191.25,185) {9};
\node[right, align=left]
at (axis cs:236.25,125) {10};

\end{axis}
\end{tikzpicture}%
}

  \subcaption{Full coordination}
  \label{fig: Example1_3_1}
\end{minipage}
\begin{minipage}{0.48\columnwidth}
  \centering
 {\scriptsize\begin{tikzpicture}
\begin{axis}[%
width=0.99\columnwidth,
height=0.75\columnwidth,
at={(0in,0in)},
scale only axis,
xmin=0,
xmax=260,
xtick={0,50,100,150,200,250},
xticklabels=\empty,
ymin=0,
ymax=200,
yticklabels=\empty,
axis background/.style={fill=white},
]
\addplot [color=black, line width=2.3pt]
  table[row sep=crcr]{%
100	90\\
20	10\\
};

\addplot [color=black, line width=2.3pt]
  table[row sep=crcr]{%
100	90\\
45	135\\
};

\addplot [color=black, line width=2.3pt]
  table[row sep=crcr]{%
175	90\\
200	20\\
};

\addplot [color=black, line width=2.3pt]
  table[row sep=crcr]{%
100	90\\
120	55\\
};

\addplot [color=black, line width=2.3pt]
  table[row sep=crcr]{%
140	125\\
165	140\\
};

\addplot [color=black, line width=2.3pt]
  table[row sep=crcr]{%
175	90\\
135	85\\
};

\addplot [color=black, line width=2.3pt]
  table[row sep=crcr]{%
100	90\\
98	119\\
};

\addplot [color=black, line width=2.3pt]
  table[row sep=crcr]{%
200	150\\
220	107\\
};

\addplot [color=black, line width=2.3pt]
  table[row sep=crcr]{%
200	150\\
185	185\\
};

\addplot [color=black, line width=2.3pt]
  table[row sep=crcr]{%
200	150\\
230	125\\
};

\addplot [color=green, line width=0.9pt]
  table[row sep=crcr]{%
100	90\\
20	10\\
};

\addplot [color=green, line width=0.9pt]
  table[row sep=crcr]{%
100	90\\
45	135\\
};

\addplot [color=green, line width=0.9pt]
  table[row sep=crcr]{%
140	125\\
165	140\\
};

\addplot [color=green, line width=0.9pt]
  table[row sep=crcr]{%
100	90\\
120	55\\
};

\addplot [color=green, line width=0.9pt]
  table[row sep=crcr]{%
175	90\\
135	85\\
};

\addplot [color=green, line width=0.9pt]
  table[row sep=crcr]{%
175	90\\
200	20\\
};

\addplot [color=green, line width=0.9pt]
  table[row sep=crcr]{%
100	90\\
98	119\\
};

\addplot [color=green, line width=0.9pt]
  table[row sep=crcr]{%
200	150\\
220	107\\
};

\addplot [color=green, line width=0.9pt]
  table[row sep=crcr]{%
200	150\\
185	185\\
};

\addplot [color=green, line width=0.9pt]
  table[row sep=crcr]{%
200	150\\
230	125\\
};

\addplot [color=green, line width=0.9pt]
  table[row sep=crcr]{%
120 55\\
175 90\\
};

\addplot [color=green, line width=0.9pt]
  table[row sep=crcr]{%
140 125\\
185 185\\
};

\addplot [color=green, line width=0.9pt]
  table[row sep=crcr]{%
200 150\\
200 20\\
};

\addplot [color=green, line width=0.9pt]
  table[row sep=crcr]{%
100 90\\
135 85\\
};

\node[font=\bfseries,draw=blue,rectangle,fill=blue,minimum size=3mm,font=\color{white}] at (100,90){\textbf{1}};
\node[font=\bfseries,draw=blue,rectangle,fill=blue,minimum size=3mm,font=\color{white}] at (175,90){\textbf{2}};
\node[font=\bfseries,draw=red,rectangle,fill=red,minimum size=3mm,font=\color{white}] at (140,125){\textbf{3}};
\node[font=\bfseries,draw=red,rectangle,fill=red,minimum size=3mm,font=\color{white}] at (200,150){\textbf{4}};

\addplot [color=blue, draw=none, mark size=3.0pt, mark=*, mark options={solid, fill=blue, blue}]
  table[row sep=crcr]{%
20	10\\
45	135\\
165	140\\
200	20\\
120	55\\
};

\addplot [color=red, draw=none, mark size=3.0pt, mark=*, mark options={solid, fill=red, red}]
  table[row sep=crcr]{%
98	119\\
135	85\\
220	107\\
185	185\\
230	125\\
};

\node[right, align=left]
at (axis cs:26.25,10) {1};
\node[right, align=left]
at (axis cs:51.25,135) {2};
\node[right, align=left]
at (axis cs:171.25,140) {3};
\node[right, align=left]
at (axis cs:206.25,20) {4};
\node[right, align=left]
at (axis cs:126.25,55) {5};
\node[right, align=left]
at (axis cs:99,130) {6};
\node[right, align=left]
at (axis cs:141.25,91) {7};
\node[right, align=left]
at (axis cs:226.25,107) {8};
\node[right, align=left]
at (axis cs:191.25,185) {9};
\node[right, align=left]
at (axis cs:236.25,125) {10};

\end{axis}
\end{tikzpicture}%
}

  \subcaption{R,Optimal}
  \label{fig: Example1_4_1}
\end{minipage}


\caption{Illustration of the association and coordination. Topology is identical to that of Fig.~\ref{fig: ToyExample}. A black (similarly green) line from BS $b$ to UE $u$ indicates that $[\bA]_{bu}=1$ (similarly $[\bC]_{bu}=1$). In \subref{fig: Example1_1_1}, every BS estimates only the channel of its associated UEs. Setting of \subref{fig: Example1_2_1} is identical to that of \subref{fig: Example1_1_1} except an extra coordination $[\bC]_{16} = 1$ to reduce inter-operator interference of UE 1. In \subref{fig: Example1_3_1}, every BS estimates the channel of every UEs. \subref{fig: Example1_4_1} shows the optimal association and coordination for $(N_{\BS}, N_{\UE}) = (8, 2)$, obtained from a variant $\calP_1$ with national roaming with $P_z^{\max} = 120$.}
\label{fig: Example1_1}
\end{figure}

Table~\ref{table: Example1} shows the performance of the network under three scenarios of Fig.~\ref{fig: Example1_1} and the optimal solution, obtained from $\calP_1$ and its national roaming variation. From this table, coordination substantially reduces the interference and improves both the network sum rate and the minimum UE rate. This improvement is significant for UE 6, which is served by BS 3 (belongs to the red operator) but is located very close to BS 1 (belongs to the blue operator).
Imposing $[\bC]_{16} = 1$ leads to a substantial reduction of $I^{(3)}$ and thus to an improvement in the achievable rate.

For the small antenna setting $(N_{\BS} = 8, N_{\UE} = 2)$ and for the topology of this example, the availability of national roaming can substantially reduce the overall coordination overhead by selecting a much better association. The optimal serving BS for UE 6 is now BS 1, and consequently, the coordination cost reduces from 100 (i.e., inter-operator cost) to 1 (i.e., $p_b$ for associated UEs). The use of large antenna arrays $(N_{\BS} = 64, N_{\UE} = 16)$ reduces the interference footprint and the need for coordination. Still, selecting a better association and coordination solution lead to an improvement in the rate performance. However, as mentioned before, this may entail a formidable signaling overhead.

\begin{table*}[t]
\centering
\caption{Performance of association and coordination of Fig.~\ref{fig: Example1_1}. O1 and O2 stand for operator 1 (blue) and operator 2 (red). Rates are in Gbps. The table shows the average of the various interference terms that UE 6 observes, namely $I1 := I_{b6}^{(1)}/{\rho^{\mathrm{Rx}}_{b6}}$, $I2 := I_{b6}^{(2)}/{\rho^{\mathrm{Rx}}_{b6}}$, and $I3 := I_{b6}^{(3)}/{\rho^{\mathrm{Rx}}_{b6}}$. Normalized rate of UE 6 shows the rate improvement with respect to scenario (a), baseline,  with the same number of antennas. Rate of UE 6 is 0.301 Gbps with $(N_{\BS}=8, N_{\UE}=2)$, and 1.884 Gbps with $(N_{\BS}=64, N_{\UE}=16)$. ``Optimal,$x$'' corresponds to the solution of $\calP_1$ with the coordination budget $P_z^{\max} = x$. ``R,Optimal,$x$'' corresponds to the national roaming variant of $\calP_1$ with $P_z^{\max} = x$.}
\label{table: Example1}
\renewcommand{\tabcolsep}{3pt}
\renewcommand{\arraystretch}{1.1}
\scriptsize{
\begin{tabular}{c|cccccc}
\toprule
\multirow{2}{*}{\# Antennas}  & \multirow{2}{*}{Scenario} & \multirow{2}{*}{\begin{tabular}[c]{@{}c@{}}Sum rates of UEs \\ $[{\mbox{O1}}, {\mbox{O2}}]$ \end{tabular}}& \multirow{2}{*}{\begin{tabular}[c]{@{}c@{}}Min rate of UEs \\ $[{\mbox{O1}}, {\mbox{O2}}] $ \end{tabular}} &  \multirow{2}{*}{\begin{tabular}[c]{@{}c@{}}Average normalized interference \\ $\left[\mathbb{E}\left[I1\right], \mathbb{E}\left[I2\right], \mathbb{E}\left[I3\right] \right]$ \end{tabular}} & \multirow{2}{*}{\begin{tabular}[c]{@{}c@{}}Rate improvement \\ of UE 6 (\%)\end{tabular}} & \multirow{2}{*}{\begin{tabular}[c]{@{}c@{}}Coordination cost \\ $[{\mbox{O1}}, {\mbox{O2}}] $ \end{tabular}}  \\ \\ \midrule
\multirow{5}{*}{\begin{tabular}[c]{@{}c@{}}$N_{\BS} = 8$\\ $ N_{\UE} = 2$\end{tabular}} & \subref{fig: Example1_1_1} & $\begin{bmatrix} 2.120& 2.463 \end{bmatrix}$ & $\begin{bmatrix} 0.247& 0.301 \end{bmatrix}$ & $\begin{bmatrix} 0.000& 0.098 & 3.044 \end{bmatrix} $ & 0 & $\begin{bmatrix} 5 & 5 \end{bmatrix}$ \\
& \subref{fig: Example1_2_1} & $\begin{bmatrix} 1.968& 2.896 \end{bmatrix}$ & $\begin{bmatrix} 0.222& 0.440 \end{bmatrix}$ & $\begin{bmatrix} 0.000& 0.087 & 0.247 \end{bmatrix} $ & 148 & $\begin{bmatrix} 5 & 105 \end{bmatrix}$ \\
& \subref{fig: Example1_3_1} & $\begin{bmatrix} 4.591& 6.297 \end{bmatrix}$ & $\begin{bmatrix} 0.710& 1.180 \end{bmatrix}$ & $\begin{bmatrix} 0.000& 0.068 & 0.240 \end{bmatrix} $ & 346 & $\begin{bmatrix} 1055 & 1055 \end{bmatrix} $\\
& Optimal,120 & $\begin{bmatrix} 2.534& 3.126 \end{bmatrix}$ & $\begin{bmatrix} 0.327& 0.518 \end{bmatrix}$ & $\begin{bmatrix} 0.000& 0.071 & 0.246 \end{bmatrix} $ & 156 & $\begin{bmatrix} 115 & 115 \end{bmatrix} $ \\
& R,Optimal,120 & $\begin{bmatrix} 4.851& 4.337 \end{bmatrix}$ & $\begin{bmatrix} 0.652& 1.105 \end{bmatrix}$ & $\begin{bmatrix} 0.000& 0.070 & 0.240 \end{bmatrix} $ & 245 & $\begin{bmatrix} 116 & 114 \end{bmatrix} $ \\
\midrule
\multirow{5}{*}{\begin{tabular}[c]{@{}c@{}}$N_{\BS} = 64$\\ $ N_{\UE} = 16$\end{tabular}} & \subref{fig: Example1_1_1} & $\begin{bmatrix} 10.476& 11.393 \end{bmatrix}$ & $\begin{bmatrix} 1.854& 1.884 \end{bmatrix}$ & $\begin{bmatrix} 0.000 & 0.001 & 0.007 \end{bmatrix} $ & 0 & $\begin{bmatrix} 5 & 5 \end{bmatrix}$ \\
& \subref{fig: Example1_2_1} & $\begin{bmatrix} 10.477& 11.886 \end{bmatrix}$ & $\begin{bmatrix} 1.912& 2.223 \end{bmatrix}$ & $\begin{bmatrix} 0.000& 0.001 & 0.002 \end{bmatrix} $ & 25 & $\begin{bmatrix} 5 & 105 \end{bmatrix}$ \\
& \subref{fig: Example1_3_1} & $\begin{bmatrix} 13.733& 15.387 \end{bmatrix}$ & $\begin{bmatrix} 2.642& 3.018 \end{bmatrix}$ & $\begin{bmatrix} 0.000& 0.000 & 0.000 \end{bmatrix} $ & 65 & $\begin{bmatrix} 1055 & 1055 \end{bmatrix} $ \\
& Optimal,120 & $\begin{bmatrix} 12.921& 13.968 \end{bmatrix}$ & $\begin{bmatrix} 2.483& 2.709 \end{bmatrix}$ & $\begin{bmatrix} 0.000& 0.001 & 0.001 \end{bmatrix} $ & 47 & $\begin{bmatrix} 115 & 115 \end{bmatrix} $ \\
& Optimal,1055 & $\begin{bmatrix} 14.263& 15.908 \end{bmatrix}$ & $\begin{bmatrix} 2.651& 2.966 \end{bmatrix}$ & $\begin{bmatrix} 0.000& 0.000 & 0.000 \end{bmatrix} $ & 68 & $\begin{bmatrix} 1055 & 1055 \end{bmatrix} $ \\
\bottomrule
\end{tabular}
}
\end{table*}

\section{Hybrid Solution Approach}\label{sec: hybrid-driven-solution}
So far, we have observed that neither $\calP_1$ nor its distributed variant can be solved in practice due to missing CSI and lack of proper rate models.
Data-driven approaches bypass the need for precise modeling techniques and are thereby
less sensitive to missing features and modeling inaccuracies.
In this section, we propose that the learning task continuously refines the rate model of every UE rather than optimizing the decision variables. The model-based part then uses the updated rate models to find proper association and coordination strategies.

To enable this hybrid solution approach, we introduce two types of frames, training and operation, designed to improve the interplay among the exploration and exploitation and quality of service at \acp{UE}. In the \textit{training frames}, the BSs and UEs use a randomized policy to explore the space of ``proper'' solutions for $(\bA,\bC)$, formally described in Section~\ref{sec: results11}, and to improve the rate models. In the \textit{operation frames}, the operators apply a previously found good solution to protect the UE performance from potentially weak rates of some candidate $(\bA,\bC)$. The new solutions will be applied to the operation frames only after passing a predefined confidence on their rate performance, measured in several training frames. Fig.~\ref{fig: ProposedApproach} illustrates the proposed hybrid approach.
\begin{figure}[!t]
  \centering
 {\scriptsize\begin{tikzpicture}[scale=1, auto, >=stealth']
    \matrix[ampersand replacement=\&, row sep=4mm, column sep=5mm] {
      \node[draw,rectangle,thick,minimum height=2em,minimum width=2em] (F1) {Channel estimation}; \&
      \node[draw,rectangle,thick,minimum height=2em,minimum width=2em] (F2) {Precoding design}; \&
      \node[draw,rectangle,thick,minimum height=2em,minimum width=2em] (F3) {Rate measurements}; \& \\
      \& \\ \\
      \node[draw,rectangle,thick,minimum height=2em,minimum width=2em,fill=green!20!white] (F5) {Rate models update};  \\
};
    \draw [->,thick] (F1) -- (F2);
    \draw [->,thick] (F2) -- (F3);
    \draw [->,thick,rounded corners] ($(F3.south west)!0.66!(F3.south east)$)  |-  (F5.east) node[pos=0.7,below]{every training frame};
    \draw [->,thick] ($(F5.north west)!0.33!(F5.north east)$)  --  ($(F1.south west)!0.33!(F1.south east)$);
    \draw [->,thick,rounded corners] ($(F3.south west)!0.33!(F3.south east)$) node[pos=0.1,below]{every CI} to [ncbar=2em] ($(F1.south west)!0.66!(F1.south east)$);
%
\end{tikzpicture} }

\caption{Illustration of our hybrid spectrum sharing approach. White boxes represent the model-based part, and green box is for the data-driven part.}
\label{fig: ProposedApproach}
\end{figure}
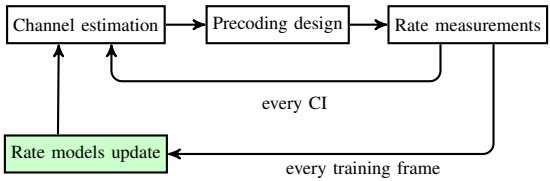

\subsection{Data-driven Part}\label{sec: data-driven-part}
Developing a solution approach for $\calP_1$ is challenging. First, due to the lack of a closed-form solution, we need iterative approaches to solve $\calP_1$. These solvers must evaluate the objective function for several $\bA$ and $\bC$ {matrices, until} convergence. Thus, one needs to send additional pilots to evaluate {the updated} combining vectors at the UEs
(which change as $\bA$ is updated), and estimate some new channels $\{\bH_{bu}\}$ for some $b$ and $u$.
These additional pilot transmissions and channel estimations can be very expensive
as we may need many iterations before convergence, and we may typically end up in a situation
where we have to estimate almost all the channels; clearly this is impractical in a cellular network.
Moreover, it is at odds with the coordination cost model~\eqref{eq: coordination-cost}, where we consider the cost associated with the final solution only.
Second, when we know the effective channels corresponding to the final solution, every BS computes $\rho^{\mathrm{Rx}}_{bu}$ and $I_{bu}^{(1)}$ from \eqref{eq: Rx-power} and \eqref{eq: intra-cell-interference-digital-precoder}, and feed them back to the cloud server.
However, we have access only to some summands of $I_{bu}^{(2)}$ and $I_{bu}^{(3)}$ for which the respective entry of $\bC$ is 1. Consequently, the central controller cannot compute $I_{bu}^{(2)}+I_{bu}^{(3)}$ and therefore the objective function.

To address these challenges, the data-driven part takes as \textit{input} the network topology, the association matrix $\bA$, the coordination matrix $\bC$, the effective channels $\overline{\bH}_b$, and \textit{outputs} an approximation of the rate of UE $u$, denoted by $\widehat{r}_{u}$. More specifically, the data-driven part is comprised of two components: a dataset and a learning method. Each entry of the dataset includes $(\bA,\bC,\overline{\bH}_b$,$\{r_{u}\}_{u \in \calU})$, while the learning method approximates the rate function. We maintain a dataset at the cloud server and update it before and after every training frame; see Section~\ref{section: training-frames}.

At every \ac{CI}, BS $b$ measures $r_{u}$ for its associated UEs (having $[\bA]_{bu} = 1$). This is simply done by a feedback from the UE reporting its throughput in this CI.
It collects these values and reports them to the cloud server prior to every training frame. The server updates the input-output dataset along with the mapping $\widehat{r}_{u}(\bA, \bC)$ for all $u\in\calU$, and computes the next tuple $(\bA,\bC)$ to be examined in the following training frame. This is done by the \call{Explore} function. After that frame, the cloud updates the dataset and the rate models and decide whether to apply new association and coordination solutions to the subsequent operation frames.

In Section~\ref{sec: initializations}, we discuss how to initialize the rate models. The cloud server then gradually updates these models with any new entry in the dataset through the \call{Update} procedure. The other functions of this algorithm, called by the operators, will be illustrated in Algorithm~\ref{alg: Cloud}.

\begin{algorithm}[t]
\caption{\small Cloud Server}
\label{alg: Cloud}
\begin{algorithmic}[1]
\small
\Procedure{\callf{Update}}{$\bA_0$,$\bC_0$,$\{r_{u}\}$}
 \State Amend new entry $(\bA_0,\bC_0,\{r_{u}\})$ to the dataset
 \State Update the rate functions $\{\widehat{r}_{u}\}$
\EndProcedure
\vspace{3mm}
\Function{\callf{Initialize}}{$\{L_{bu}\}$ if available}
  \State \Return $\bA^{(0)}$ and $\bC^{(0)}$, as described in Section~\ref{sec: Model-based-part}
\EndFunction
\vspace{3mm}
\Function{\callf{Download}}{}
  \State \Return $\{\widehat{r}_{u}\}_{u}$ for all UE $u$
\EndFunction
\vspace{3mm}

\Function{\callf{Optimize}}{$\bA^{(0)}$, $\bC^{(0)}$} 
\State Initialize $\bA^{(0)}$, $\bC^{(0)}$ (described in Section~\ref{sec: Model-based-part})
\For{$k=1,2,3,\ldots$}
 \State Run A-step and find $\bA^{(k+1)}$ using \eqref{eq: A-step}
 \State Run C-step and find $\bC^{(k+1)}$ using \eqref{eq: C-step}
 \If {Convergence criteria met}
 \State Set $\bA^\star \leftarrow \bA^{k}$ and $\bC^\star \leftarrow \bC^{k}$, and break the loop
 \EndIf
\EndFor
 \State \Return $\bA^\star$ and $\bC^\star$
\EndFunction
\vspace{3mm}
\Function{\callf{Explore}}{$\bA^{\star}$, $\bC^{\star}$, $\calF$}
\State Set
\vspace{-2mm}
\begin{equation*}
\left(\bA^{\text{tf}}, \bC^{\text{tf}}\right) \leftarrow
\begin{cases}
  \text{a random}~ \left(\bA, \bC\right) \in \calF, & \mbox{with probability}~\epsilon \\
  \left(\bA^\star, \bC^\star\right), & \mbox{otherwise}.
\end{cases}
\end{equation*}
\State \Return $\bA^{\text{tf}}$ and $ \bC^{\text{tf}}$
\EndFunction
\end{algorithmic}
\vspace{-1mm}
\end{algorithm}
\vspace{-4mm}

\subsection{Model-based Part}\label{sec: Model-based-part}
Given the updated rate models, the cloud server formulates and solves an optimization problem similar to $\calP_1$ and finds the new association and coordination solutions. In the following, we derive the modified optimization problem and develop a solution algorithm.

We start by re-writing the optimization problem as a function of $\bA$ and $\bC$.
We write \eqref{eq: coordination-cost} as
\begin{equation}\label{eq: temp}
\bP = \bP_0 + \bA\left(p_b \mathbf{1}- \bP_0\right)
\end{equation}
where $\mathbf{1}$ is a matrix of ones having appropriate size.
Then, we can rewrite the coordination cost~\eqref{eq: coordination-cost} as
\begin{align*}
\sum_{u\in\calU_z}\sum_{b\in\calB}[\bC]_{bu} [\bP]_{bu} & =
\sum_{u\in\calU_z}\sum_{b\in\calB} \left[ \bC \circ \bP \right]_{bu}  = \sum_{u\in\calU_z}\left[\bP^{\trans} \bC\right]_{uu} \nonumber \\
& \overset{\eqref{eq: temp}}{=}
\sum_{u\in\calU_z}\left[\left(\bP_0 + \bA\left(p_b \mathbf{1} - \bP_0\right)\right)^{\trans} \bC\right]_{uu}
\end{align*}
where $\circ$ is the Hadamard product, and $(\cdot)^{\trans}$ is the transpose operation.
If required, every operator can obtain an approximation of the rate functions of its UEs through the \call{Download} function of Algorithm~\ref{alg: Cloud}, and then find an approximation of $f_z(\bA,\bC)$, denoted by $\widehat{f}_z(\bA,\bC)$, for any $\bA$ and $\bC$, where $\widehat{f}_z = \sum_{u \in \calU_z}\, \log \widehat{r}_{u} $.
We can now write {the} modified optimization problem as:
\begin{subequations}\label{eq: optimal-spectrum-sharing-m}
\begin{alignat}{3}
\label{eq: objective-P1-m}
\calP_{1R}\hspace{-0.8mm}:~\hspace{-1.3mm} \max_{\bA, \bC} & \hspace{1.5mm} && \sum\nolimits_{z=1}^{Z}\alpha_z \widehat{f}_z (\bA,\bC) \:,
\\
\label{eq: const-unique-association-m}
{\text{s.t.}}& && {\text{Constraints \eqref{eq: const-unique-association}, \eqref{eq: cell-size-p1}, \eqref{eq: binary-variables-11}, and \eqref{eq: binary-variables}}}
\\
\label{eq:MaxCommBudget}
& &&
\sum\nolimits_{u\in\calU_z}\left[\left(\bP_0 + \bA\left(p_b \mathbf{1} - \bP_0\right)\right)^{\trans} \bC\right]_{uu}  \leq P_{z}^{\max} , \nonumber \\[-1mm]
& && \hspace{50mm} \forall 1 \leq z \leq Z \:,
\end{alignat}
\end{subequations}
Notice that the computational complexity of \eqref{eq: optimal-spectrum-sharing-m} is of the same order of magnitude as that of \eqref{eq: optimal-spectrum-sharing}, and we can reuse the existing solution algorithms of the pure model-based approach, \eqref{eq: optimal-spectrum-sharing-m}, in the model-based part of our hybrid approach.
However, the main benefit of \eqref{eq: optimal-spectrum-sharing-m} is having a much lower signaling complexity and latency to acquire the needed channel state information. In many cases, we may not be able to compute the objective function of \eqref{eq: optimal-spectrum-sharing} due to the heavy signaling complexity and other challenges involved; see Section~\ref{sec: practical_considerations}.

In general, the objective $\widehat{f}_z$ is not jointly convex in $\bA$ and $\bC$, and the space of the problem is combinatorial.
Thus, we employ the block-coordinate descent (BCD) framework (also known as alternating optimization)~\cite{Razaviyayn_BCD_12}, where $\calP_{1R}$ is split into two subproblems solved iteratively: (A-step) to find the optimal association and (C-step) to find the optimal coordination.

Denoting by $\bA^{(k)}$ and $\bC^{(k)}$ denote the values for $\bA$ and $\bC$ at iteration $k$, {BCD} yields the following update rules:
\begin{subequations}\label{eq: A-step}
\begin{alignat}{3}
\label{eq: objective-P1-m2} 
{\text{\underline{(A-step)}: }} \hspace{12mm} & && \nonumber \\
\bA^{(k+1)} \in \underset{\bA}{\mathrm{argmax}} & \hspace{1.2mm} && \sum\nolimits_{z=1}^{Z}\alpha_z \widehat{f}_z (\bA,\bC^{(k)}) \:,
\\
{\text{s.t.}}& && {\text{Constraints \eqref{eq: const-unique-association}, \eqref{eq: cell-size-p1}, and \eqref{eq: binary-variables-11}}}
\\
& &&
\sum_{u\in\calU_z}\left[\left(\bP_0 + \bA\left(p_b \mathbf{1}- \bP_0\right)\right)^{\trans} \bC^{(k)}\right]_{uu} \nonumber \\[-1mm]
& && \hspace{24mm} \leq  P_{z}^{\max}\:, \hspace{2mm} \forall 1 \leq z \leq Z ,
\\
& &&
\label{eq: const-binary}
\left[\bA\right]_{b u}  \in \{0,1 \} \:, \hspace{2mm} \forall b \in \calB, u \in \calU \:.
\end{alignat}
\end{subequations}

\begin{subequations}\label{eq: C-step}
\begin{alignat}{3}
{\text{\underline{(C-step)}: }} \hspace{12mm} & && \nonumber \\
\bC^{(k+1)} \in \underset{\bC}{\mathrm{argmax}} & \hspace{1.2mm} && \sum_{z=1}^{Z}\alpha_z \widehat{f}_z (\bA^{(k+1)},\bC) \:,
\\
{\text{s.t.}}& &&
\sum_{u\in\calU_z}\left[\left(\bP_0 + \bA^{(k+1)}\left(p_b \mathbf{1} - \bP_0\right)\right)^{\trans} \bC\right]_{uu} \nonumber\\[-1mm]
& && \hspace{24mm} \leq P_{z}^{\max}\:, \hspace{2mm}  \forall 1 \leq z \leq Z \:, \\
& &&
\left[\bC\right]_{b u}  \in \{0,1 \} \:, \hspace{2mm} \forall b \in \calB, u \in \calU \:.
\end{alignat}
\end{subequations}

Although the above subproblems are combinatorial, they may be still be solved effectively using binary programming or branch-and-bound solvers~\cite{Berstakas_nonlinear_99}. We must emphasize that the use of BCD drastically reduces the size of the search space from $\calO( 2^{ |\calB|^2 |\calU|^2 } )$, for the joint optimization optimization in $\calP_{1R}$, to $\calO( 2^{ |\calB| |\calU| } )$ for each BCD iteration. Moreover, we can further seek sufficient conditions on the approximation functions $\widehat{f}_{u}$. For instance, when the learning function is bilinear in $\bA$ and $\bC$, and the coordination penalty matrix consists of integers values, then linear program relaxation of these sub-problems is optimal or close to optimal~\cite{Berstakas_nonlinear_99}. In the future, we will investigate efficient solution methods and relaxations for $\calP_{1R}$. This current work, however, is aimed at showing the usefulness of this approach, rather than its large-scale implementation.

Moreover, not being able to show the local optimality is a known downside of almost all first-order methods (including BCD) in a nonconvex landscape. Indeed, the iterative algorithms may converge to a saddle point, which is stationary but neither local maxima nor minima. However, recent studies showed that the gradient noise in the stochastic (mini-batch) gradient along and the use of the perturbed gradient descent method, as we have used in our work, are efficient approaches to escape first-order saddle points~\cite{jin2017escape}.


Let $\bA^{\text{of}}$ and $\bC^{\text{of}}$ denote the association and coordination matrices for operation frames, $\bA^{\text{tf}}$ and $\bC^{\text{tf}}$ denote the association and coordination matrices for a training frame, and $\bA^{(k)}$ and $\bC^{(k)}$ denote the association and coordination matrices at iteration $k$ of BCD.
Algorithm~\ref{alg: MainFlowchart-hybrid} is a pseudo-code of our hybrid solution approach. Below, we show the monotonically increasing nature of the BCD updates.
\begin{algorithm}[!t]
\caption{\small Hybrid Model-based and Data-driven Spectrum Sharing}
\label{alg: MainFlowchart-hybrid}
\begin{algorithmic}[1]
\small
\Require
      \ac{CI} index $n$; An indexed sequence of training and operation frames; a feasibility space for the association and coordination $\calF$
\Ensure
Beamforming vectors in every \ac{CI}, optimal $\bA$ and $\bC$
\State Run $(\bA^{\text{of}},\bC^{\text{of}})=\callf{Initialize}()$ at the cloud
\State Set $\bA \leftarrow \bA^{\text{of}}$ and $\bC \leftarrow \bC^{\text{of}}$
\For{$n=1,2,3,\ldots$}
\State Every BS $b$ estimates $\bH_{bu}$ for all $\{u\mid [\bA]_{bu}=1\}$
\State Every BS $b$ designs $\bw_{u}^{\UE}$ based on \eqref{eq: AnalogCombiner} for its associated UEs
\State Every BS $b$ estimates $(\bw_{u}^{\UE})^{\herm}\bH_{bu}$ for all $\{u\mid [\bC]_{bu}=1\}$
\State Find the precoding vectors from~\eqref{eq: DigitalBeamforming}
\State Operate with those precoding and combining vectors
\State Measure $r_{u}$ at the end of \ac{CI} $n$ and record it
\If{\ac{CI} $n$ is a \emph{training frame}}
\State Run $\callf{Update}(\bA,\bC,\{r_{u}\})$ at the cloud for rates obtained from all UEs in \ac{CI} $n$
\State Set $\bA^{(0)}\leftarrow \bA^{\text{of}},\quad \bC^{(0)}\leftarrow\bC^{\text{of}}$
\State\label{alg: BCD-step} Run $(\bA^\star, \bC^\star) = \callf{Optimize}(\bA^{(0)}, \bC^{(0)})$ at the cloud
\State Run $(\bA^{\text{new}}, \bC^{\text{new}}) = \callf{Explore}(\bA^\star, \bC^\star,\calF)$
\State Clear recorded rates at every BS
\If{\textit{confidence criteria} met for $(\bA^{\text{new}}, \bC^{\text{new}})$}
\State Set $\bA^{\text{of}} \leftarrow \bA^{\text{new}}$ and $\bC^{\text{of}} \leftarrow \bC^{\text{new}}$
\EndIf
\State Set $\bA \leftarrow \bA^{\text{of}}$ and $\bC \leftarrow \bC^{\text{of}}$
\EndIf
\If{\ac{CI} $(n + 1)$ is a \emph{training frame}}
\State Run $\callf{Update}(\bA,\bC,\{r_{u}\})$ at the cloud for rates obtained from all UEs in the previous operation frames
\State Set $\bA^{(0)}\leftarrow \bA^{\text{of}},\quad \bC^{(0)}\leftarrow\bC^{\text{of}}$
\State \label{alg: BCD-step2} Run $(\bA^\star, \bC^\star) = \callf{Optimize}(\bA^{(0)}, \bC^{(0)})$ at the cloud
\State Run $(\bA^{\text{tf}}, \bC^{\text{tf}}) = \callf{Explore}(\bA^\star, \bC^\star,\calF)$
\State Set $\bA \leftarrow \bA^{\text{tf}}$ and $\bC \leftarrow \bC^{\text{tf}}$
\State Clear recorded rates at every BS
\EndIf
\EndFor
\end{algorithmic}
\end{algorithm}

\begin{lemma}[Convergence of BCD]\label{lemma: BCD-convergence}
Let $\widehat{f}_z$ be continuous biconcave in $\bA$ and $\bC$.
Then, the BCD updates in~\eqref{eq: A-step} and~\eqref{eq: C-step} satisfy $\widehat{f}_z (\bA^{(k)},\bC^{(k)}) \leq \widehat{f}_z (\bA^{(k+1)},\bC^{(k)}) \leq \widehat{f}_z (\bA^{(k+1)},\bC^{(k+1)})$.
Moreover, the updates converge to a limit point $\lim_{k \rightarrow \infty}~ \widehat{f}_z (\bA^{(k)},\bC^{(k)}) $.
\end{lemma}
Although the convergence of BCD updates to a limit point is shown using standard BCD results, establishing that the limit point is stationary with respect to $\calP_{1R}$ is more challenging. Indeed, the coupling between $\bA$ and $\bC$ in constraint~\eqref{eq:MaxCommBudget} implies that the conventional BCD convergence cannot be applied to show that $\lim_{k \rightarrow \infty}~ \widehat{f}_z (\bA^{(k)},\bC^{(k)})$ is a stationary point of $\calP_{1R}$.

\subsection{Training Frames}\label{section: training-frames}
The \call{Optimize} function of the server will be re-executed before and after every training frame. The purpose of these frames is to dynamically refine the current rate models and thereby find a better association and coordination solution. Naturally, we expect a high frequency of training frames in the first few association periods (as we assume no a priori knowledge of the network), while this frequency can be decreased as we obtain more knowledge on the rate models. In the presence of non-stationary environments, where the rate distributions are changing over time, we may need to add enough training frames to enable the tracking functionality. In Section~\ref{sec: results11}, we numerically investigate how many training frames are required to find a close-to-optimal solution after a change in the number of UEs.

Before every training frame, the server gets all the new rate measurements, updates its models, and re-executes the BCD procedure. It then runs a randomized policy on a set of feasible solutions $\calF \subseteq \{0,1\}^{|\calB|\times|\calU|} \times \{0,1\}^{|\calB|\times|\calU|}$ and returns one association and one coordination matrix to be explored in the following training frame. After this exploration, the cloud updates the rate models and checks whether there is a new ``reliable'' solution to be applied in the operation frames. This reliability can be measured in terms of some predefined confidence bounds on the objective function. The consequence of this conservative approach is protecting UEs from service interruption due to unsure $\bA$ and $\bC$.

\subsection{Initializations}\label{sec: initializations}
We underline the importance of initializing both the \callf{Update} procedure and the \callf{Optimize} function. More specifically, we discuss a ``good'' starting point to speed up learning $\{ r_u \}$, and initial solutions $\bA^{(0)}, \bC^{(0)}$ to the BCD algorithm.
\subsubsection{Rate Model}
We first observe that severe path-loss, blockage, and directionality substantially reduce the interference footprint of mmWave networks in both cellular~\cite{di2014stochastic} and ad~hoc~\cite{Shokri2015Transitional} settings. In this case, we can use the well-known Gaussian approximation for the interference by an i.i.d. realization of a Gaussian process~\cite{verdu1998multiuser}. In particular,
\begin{equation}\label{eq: hatI-2-3}
\widehat{I}_{bu}^{\, (2)}  = \sum\limits_{i \in \calB_z \setminus \{b\}}  \widehat{I}_{i,u}^{\, (4)} \:, \hspace{.3cm}
\widehat{I}_{bu}^{\, (3)}  = \sum\limits_{k=1 \hfill \atop k \neq z \hfill}^{Z} \, \sum\limits_{i \in \calB_k \setminus \{b\}} \widehat{I}_{i,u}^{\, (4)} ,
\end{equation}
where $\widehat{I}_{i,u}^{\, (4)} := \sum_{j\in\mathcal{A}_{i}}\mathbb{E}\big[\lambda_{i}\big| \left(\bw_{u}^{\UE}\right)^{\herm} \bH_{iu} \bw_{i j}^{\BS} \big|^2 \big] $ denotes the interference from unintended BS $i (\neq b)$.
We can now prove the following proposition.
\begin{prop}\label{prop: initialization}
Let $\bA$ and $\bC$ be given, $[\bA]_{bu}=1$, $N_{bu}=1$, $\theta_{b u}^{\UE}$ and $\theta_{i u}^{\UE}$ be AoA of the LoS links between UE $u$ and BSs $b$ and $i$, respectively. Let $L_{iu} = \mathbb{E}[|g_{i u n}|^2]$ for $n=1$ (LoS path). Then,
\begin{align} \label{eq: intialization-interference-models}
 \widehat{I}_{i,u}^{\, (4)} =
\begin{cases}
    N_{\BS} N_{\UE} L_{iu} \rho^{\mathrm{Tx}} \left|\sinc\left(\frac{N_{\UE} (\theta_{b u}^{\UE} - \theta_{i u}^{\UE})}{2}\right)\right| , & \hspace{-3mm} \mbox{if } [\bC]_{iu} = 0 \\
    0, & \mbox{otherwise},
\end{cases}
\end{align}
where $\sinc(x)$ is $\sin(x)/x$ for $x\neq0$ and 1 for $x=0$.
\end{prop}
Notice that \eqref{eq: intialization-interference-models} is valid for $N_{bu}=1$, namely single path between BS $b$ and UE $u$. However, we have numerically observed that \eqref{eq: intialization-interference-models} indeed leads to a very good initialization of the rate models, which could be due to the sparse scattering characteristic of the mmWave systems.

From the definition of RZF, $\widehat{I}_{bu}^{\, (1)}=0$ and $\widehat{\rho}^{\, \mathrm{Rx}}_{bu}= N_{\BS} N_{\UE} L_{bu} \rho^{\mathrm{Tx}}$ for any feasible coordination solution in which a BS obtains the CSI of its associated UEs.
Using~\eqref{eq: intialization-interference-models}, we can also simplify the
expressions of $\widehat{I}_{bu}^{\, (2)}$ and $\widehat{I}_{bu}^{\, (3)}$ in \eqref{eq: hatI-2-3}. Employing these expressions, the cloud server can initialize the rate models for every association
and coordination matrices $\bA$ and $\bC$ with one of the following three scenarios:
\begin{itemize}
  \item \emph{Full topological knowledge:} If the cloud server knows a priori $L_{iu}$, $\theta_{b u}^{\UE}$ and $\theta_{i u}^{\UE}$ for all $i,b,u$ such that $[\bA]_{bu}=1$ and $[\bC]_{bu}=0$, then it substitutes~\eqref{eq: intialization-interference-models} into \eqref{eq: hatI-2-3}, and sets $\widehat{I}_{bu}^{\, (1)}=0$ and $\widehat{\rho}^{\, \mathrm{Rx}}_{bu}= N_{\BS} N_{\UE} L_{bu} \rho^{\, \mathrm{Tx}}$.
  \item \emph{Partial topological knowledge:} If the cloud server knows a priori only $L_{iu}$ for all $i\in \calB$ and $u\in \calU$, then it substitutes  
          $\widehat{I}_{i,u}^{(4)} = N_{\BS} N_{\UE} L_{iu} \rho^{\mathrm{Tx}}$ if $ [\bC]_{iu} = 0$ and otherwise 0. Note that we have used $|\sinc(x)| \leq 1$ for all $x \in \mathds{R}$. Also set $\widehat{I}_{bu}^{\, (1)}=0$ and $\widehat{\rho}^{\, \mathrm{Rx}}_{bu}= N_{\BS} N_{\UE} L_{bu} \rho^{\mathrm{Tx}}$.
  \item \emph{No topological knowledge:} In this case, the cloud server initiates the learning process by $I_{bu}^{\, (1)}(\bA, \bC) = I_{bu}^{\, (2)}(\bA, \bC) = I_{bu}^{\, (3)}(\bA, \bC) = 0$ for all $b,u,\bA,\bC$. In this case, our initialization reduces to the well-known interference-free assumption~\cite{Athanasiou-etal-2013,Yu2016Distributed,shokri2016Spectrum}. Moreover, we set $\widehat{\rho}^{\, \mathrm{Rx}}_{bu}= N_{\BS} N_{\UE} \rho^{\mathrm{Tx}}$ for all BS and UE pairs.
\end{itemize}

After the initialization, the cloud server gradually updates the rate models with any update in the dataset through the \call{Update} procedure of Algorithm~\ref{alg: Cloud}.

\subsubsection{BCD Solver}
To initialize the BCD iterations for the very first time, we use the \call{Initialize} function (in Algorithm~\ref{alg: Cloud}) with one of the following options:
\begin{itemize}
  \item \textit{Full/partial topological knowledge available:} We use the following rule as an approximation of the strongest BS association. For every $z$ and $u\in \calU_z$, $[\bA^{(0)}]_{bu} = 1$ for $b\in\argmax_{b\in\calB_z} L_{bu}$. We then set $\bC^{(0)} = \bA^{(0)}$.
  \item \textit{No topological knowledge available:} We randomly allocate UEs to BSs within the same operator. We then set $\bC^{(0)} = \bA^{(0)}$.
\end{itemize}
In the subsequent frames, we initialize the BCD solver by the current association and coordination matrices used in the operation frames.

\subsection{Illustrative Numerical Results}\label{sec: results11}
In this section, we numerically investigate the performance of our proposed spectrum sharing approach. We use the same network as that in Table~\ref{table: Example1}, a CI of 1~ms, and two antenna configurations, \textit{small} $(N_{\BS} = 8,N_{\UE}=2)$ and \textit{large} $(N_{\BS} = 64,N_{\UE}=16)$. The network is stationary during the simulation, so that the optimal association and coordination are fixed. In this case, the optimal performance of the solutions are presented in Table~\ref{table: Example1}.

For the learning task inside the \call{Update} procedure, we use a fully-connected deep neural network with 1 input layer having $2|\calB||\calA|$ nodes, 5 hidden layers each having 20 nodes, and one output layer having $|\calU|$ nodes. We use a quadratic loss (for the regression task) and train the neural network with backpropagation, mini-batch gradient method with a mini-batch size of 10 samples~\cite{Bottou2018SIAM}, and the ADAM optimizer for adaptive step-size~\cite{kingma2014adam}. To ensure escaping the first-order saddle points, we have also slightly perturbed gradients for a few times once the iterations stall \cite{jin2017escape}.
Notice that the input layer takes a concatenation of the vectorized form of $\bA$ and $\bC$, and the output layer returns the regression results for $\{\widehat{r}_u\}_u$.\footnote{We have selected this learning model as it was easy enough to train and expressive enough to model the rate function with good accuracy. Moreover, it offers enough generalization to handle the dynamic number of BSs and UEs, as numerically verified in the extended version of the manuscript~\cite{shokri2019learhningSPECSreport}. However, these choices are not unique, and we believe that some other functional approximation and training techniques (e.g., other neural network architectures or training algorithms) may be useful as well. Recall that the main contribution of this work is to develop a hybrid approach and learning-friendly architecture for spectrum sharing in mmWave networks. A detailed comparison of the impact of various functional approximation techniques (e.g., other neural network architectures or training algorithms) is an interesting future work.}

For the \call{Initialize} function, we assume the availability of the full topological knowledge, so the location of all nodes and path-loss of all links are available to the cloud. For the \call{Explore} function, we restrict the set of feasible association by limiting the cell-size to 150 meters. This is a reasonable assumption in mmWave networks, due to severe path loss and a dense BS deployment. Moreover, we enforce that every BS should estimate the effective channel toward its associated UEs. Moreover, to improve the exploitation, we gradually decay exploration parameter $\epsilon$ by setting $\epsilon \leftarrow 0.9 \times \epsilon$ after every 1000 CIs. Finding the optimal decrement rate for $\epsilon$ or even developing a deterministic exploration policy are interesting topics for future work. We have considered two benchmarks: closest BS association and Oracle (upper bound on performance). In the first benchmark, every UE is served by the closest BS. In this case, a BS acquires CSI of only its associated UEs in every CI (so no inter-BS coordination). The Oracle benchmark shows the performance of the solution of the pure model-based approach, $\calP_1$, given also in Table~\ref{table: Example1}, in which the cloud server needs perfect CSI of all channels in the network. Although we were not able to find any state-of-the-art approaches for our problem setting, we should emphasize that their potential performance would respect our benchmarks. As we shall see, the performance of our approach is very close to that of the Oracle in most cases.

Fig.~\ref{fig: SumRate} illustrates the instantaneous network sum rate of our hybrid approach.\footnote{Extended version of this paper includes more numerical results on the scalability of our method and the performance in the presence of dynamic number of UEs~\cite{shokri2019learhningSPECSreport}.} From this figure, the envelope of the sum-rate is increasing with CI index. Interestingly, we also observe that sum-rate values converge to the Oracle, which suggests that Algorithm~\ref{alg: MainFlowchart-hybrid} is asymptotically optimal in this example.
This convergence behavior validates our earlier discussions regarding the importance of initialization for the learning function; see Section~\ref{sec: initializations}. We should emphasize that the particular propagation characteristics of mmWave networks allow for that initialization. Observe that these conclusions also hold for large antenna scenario, where the increased sum-rate is due to a reduction in interference -- which is in turn due to the increased directionality. Moreover, notice that the fluctuations in Fig.~\ref{fig: LearningPerformance} are normal due to the i.i.d. realizations of the small-scale fading in every CI and the randomness in the channel estimation error.

Fig.~\ref{fig: MinRate}  shows minimum UE rate for the same numerical setup, where the above conclusions still hold. Furthermore, the increased variance of the fluctuations is a result of looking at the minimum rate, which has inherently more randomness than the sum-rate. Surprisingly, Fig.~\ref{fig: MinRate} also reveals that Algorithm~\ref{alg: MainFlowchart-hybrid} offers good robustness and fairness (with respect to the minimum rate), although the sum-rate is the objective that is maximized.
Finally, our approach substantially outperforms the closest-BS association in terms of both the network sum-rate and the minimum rate of UEs. The gain is mainly due to 1) coordination in the small antenna regime, where the interference may be stronger, and 2) load balancing over the network in the larger antenna regime, where the interference may be less dominant.

\begin{figure}[!t]
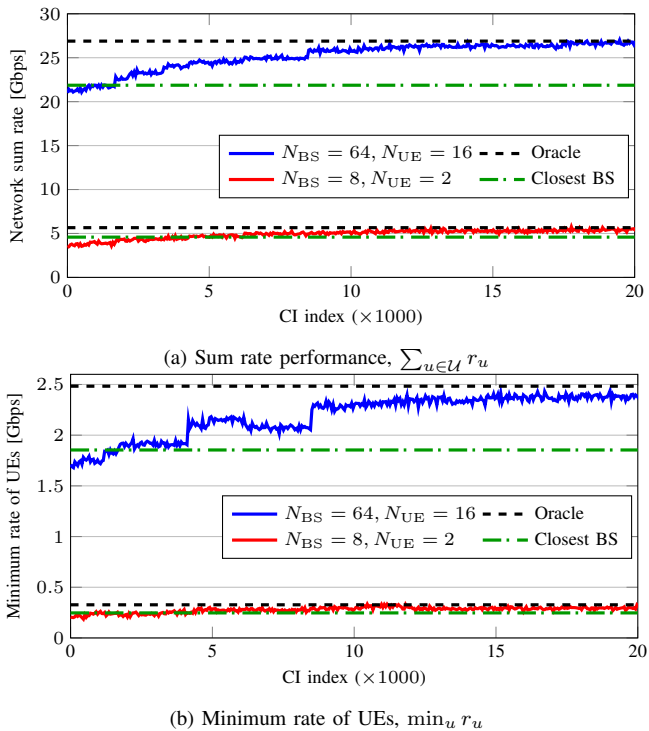

  \centering
\begin{minipage}{0.99\columnwidth}
  \centering
 {\scriptsize\input{SumRate.tex}}

  \subcaption{Sum rate performance, $\sum_{u\in\calU} r_u$}
  \label{fig: SumRate}
\end{minipage}
\begin{minipage}{0.99\columnwidth}
  \centering
 {\scriptsize\input{MinRate.tex}}

  \subcaption{Minimum rate of UEs, $\min_{u} r_u$}
  \label{fig: MinRate}
\end{minipage}
\vspace{5mm}

\caption{Illustration of the rate performance of our hybrid approach with $P_z^{\max} = 115$. The dashed black line in~\subref{fig: SumRate}, Oracle, corresponds to the solution of pure model-based approach, shown in Table~\ref{table: Example1}.}
\label{fig: LearningPerformance}
\end{figure}

We have also evaluated the performance of our approach on a much bigger network, shown in Fig.~\ref{fig: RealExampleTopology}, where each operator has 14 BSs, deployed alongside the 5th and 6th avenues of Manhattan with inter-BS distance of 75~m. Each operator has also 20 UEs in 5th Ave and 20 in 6th Ave, randomly located within the serving area. Due to the existence of many decision variables, the model-based part of our approach is a computational bottleneck in this topology. To alleviate it, we first apply a continuous relaxation to the binary constraints \eqref{eq: const-binary}, namely
\begin{equation}\label{eq: r1}
\left[\bA\right]_{b u}  \in [0,1 ] \:, \forall b \in \calB, u \in \calU\:,
\end{equation}
and then rounding to recover a binary solution. Furthermore, we assume that $[\bA]_{bu} = 1$ implies $[\bC]_{bu} = 1$ for every $b\in\calB$ and $u \in \calU$.
This is a natural simplification of the optimization problem, as a serving BS will always estimate the channels of its serving UEs.
This assumption substantially simplifies optimization problem \eqref{eq: C-step}. These simplifications, along with using a simpler interference model, allow us to scale the test network. In particular,
\begin{itemize}
\item we use the one-ball blockage model~\cite{di2014stochastic} of mmWave networks to exclude far-away transmitters from the interference model. In short, in this model, all the transmitters within a certain distance are in line-of-sights and the remaining transmitters are all blocked. We then assume infinite penetration loss. This is called the interference ball model, which has shown to be very accurate in mmWave cellular networks~\cite{Shokri2018IMSindex}. We set the distance threshold to be 150 meters, implying that for every BS-UE pair $(b,u)$ with Euclidean distance more than 150~m, $\rho^{\mathrm{Rx}}_{bu} = 0$ and consequently we set $[\bA]_{bu} = 0$.
\item We apply our interference ball model to the \call{Initialize} function of the cloud server.
\item Finally, due to the height of the buildings on the street sides, we assume that there is no signal leakage between 5th Ave and 6th Ave.
\end{itemize}
After these natural assumptions and modifications, we ran our approach on the example of Fig.~\ref{fig: RealExampleTopology}. Parameter setup for the learning tasks are the same as of Fig.~\ref{fig: LearningPerformance} except $P_z^{\max} = 220$. We have also applied the same modifications to the pure model-based Oracle, $\calP_1$, in which the cloud server has access to perfect CSI of all channels in the network.
\begin{figure}[!t]
  \centering
  \includegraphics[width=0.7\columnwidth]{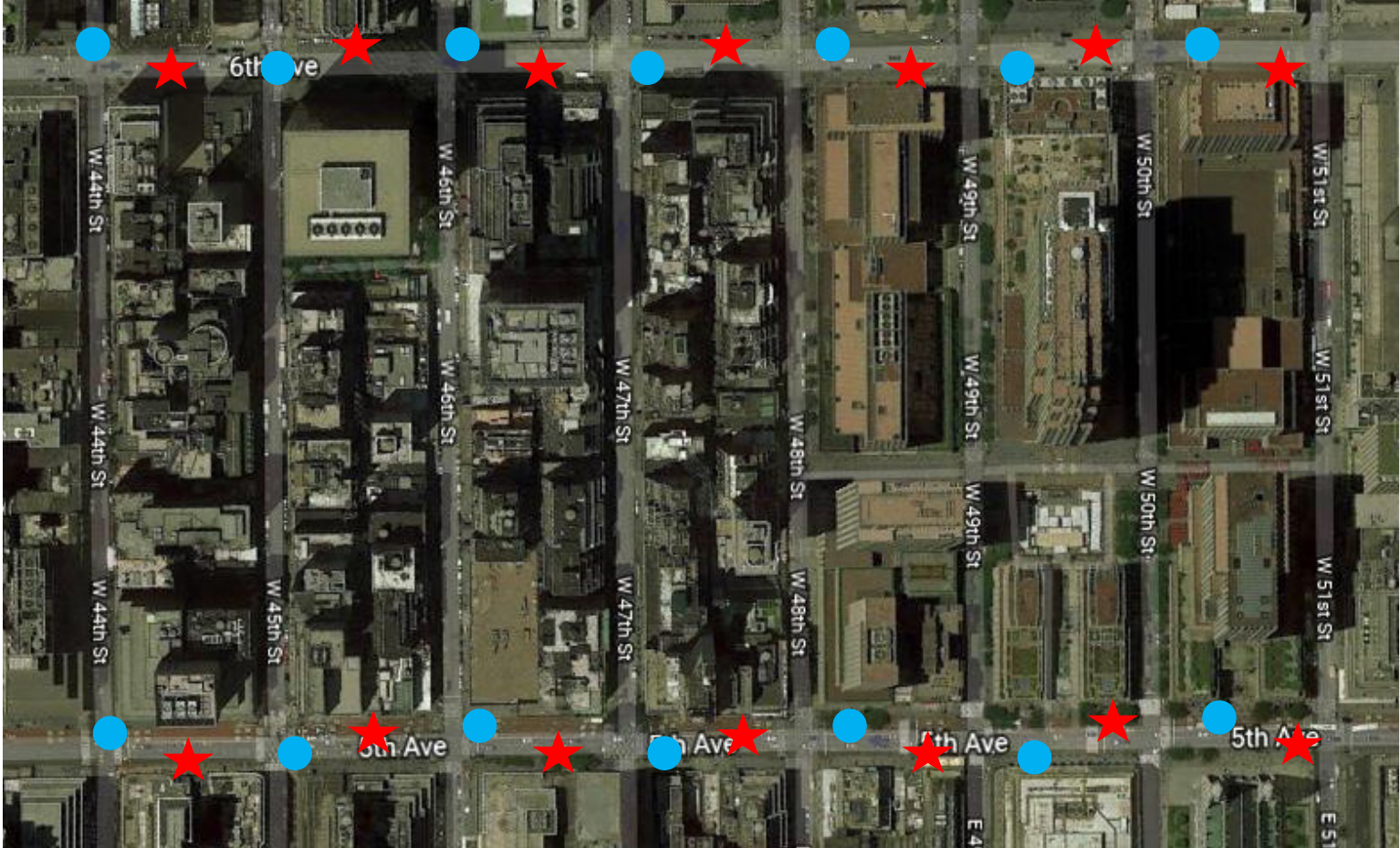}

\caption{Network topology. Stars and circles show the location of 14 BSs for blue and red operators, deployed alongside 5th and 6th avenues of Manhattan with inter-BS distance of 75~m. Each operator has 20 UEs in 5th Ave and 20 in 6th Ave, randomly located within the serving area.}
\label{fig: RealExampleTopology}
\end{figure}

Fig.~\ref{fig: RealExampleTopologySumRate} illustrates the instantaneous sum rate of the network. These curves indicate performance improvement after learning over several training frames. This convergence behavior reemphasizes our earlier discussions regarding the importance of proper initialization for the learning functions for faster convergence to the optimal solution. Moreover, our approach (which became computationally feasible for large networks due to our interference ball model) substantially outperforms the closest-BS association due to coordination and load balancing.
\begin{figure}[!t]
  \centering
 \scriptsize\input{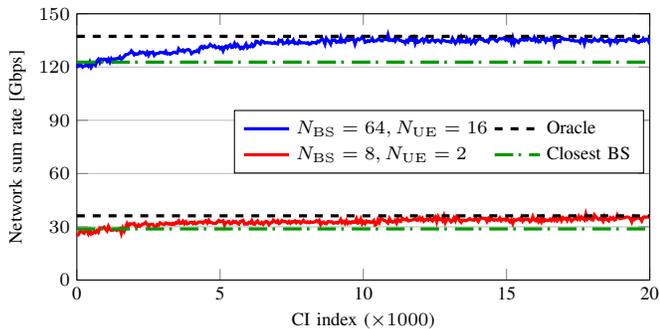}

\caption{Illustration of the rate performance of our hybrid approach with $P_z^{\max} = 220$. The Oracle corresponds to the solution of pure model-based approach.}
\label{fig: RealExampleTopologySumRate}
\end{figure}

Finally, we have evaluated the impact of a dynamic number of UEs; see Section~\ref{sec: dunamic-no-UEs} for more details on how to extend the proposed algorithm. We have considered the topology of Fig.~\ref{fig: RealExampleTopology} with 20 UEs per operator. At CI 20000 and 23000, we add one (equivalent to 5\% more UEs) and three (equivalent to 15\%) more UEs to every operator, respectively. These additional UEs are placed at random locations in 5th and 6th Avenues. Fig.~\ref{fig: DynamicNoUEs} shows the network sum rate performance. From the figure, our proposed algorithm together with our initializations can handle a minor change to the network, simulated through adding one more UE, and recover the new solution very fast, using only a few new samples. With a bigger change in the network, e.g., adding 15\% more UEs, our algorithm needs some rounds of exploration to get closer to the Oracle's performance (upper bound). In the meanwhile, thanks to our special initialization, we start from an already good solution, which gets better in time. Altogether, our algorithm can track the dynamic number of UEs and maintain the network sum-rate at a top-level.
\begin{figure}[!t]
  \centering
 \scriptsize\input{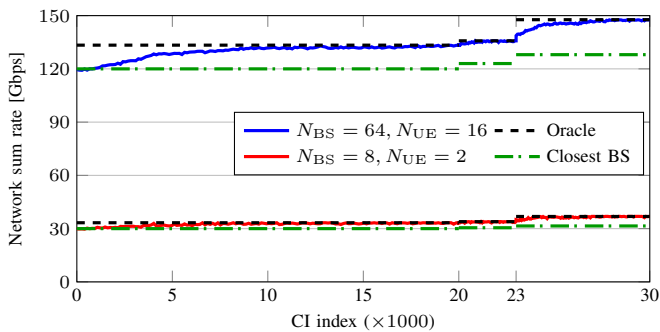}

\caption{Dynamic number of UEs for the topology of Fig.~\ref{fig: RealExampleTopology}. At CI 20000, we add one UE per operator. At CI 23000, we add 3 extra UEs per operator.}
\label{fig: DynamicNoUEs}
\end{figure}

\section{Further Discussions and Future Works}\label{sec: discussions}
\subsection{Performance in the Large Antenna Regime}
In this subsection, we evaluate the asymptotic behavior of spectrum sharing when the number of antennas grows large.
It was shown in \cite{shokri2016Spectrum} that the array response vectors at the BS and UE, i.e., $\{\ba_{\BS}(\theta)\}_{\theta}$ and $\{\ba_{\UE}(\theta)\}_{\theta}$ form an orthonormal basis, which can serve as orthogonal spatial signatures of the BSs and UEs, as $N_{\BS}$ and $N_{\UE}$ grow large.
Moreover, in this regime, there exist infinitely many spatial signatures (corresponding to different values of $\theta$). Thus, multiuser interference vanishes as a result of assigning different signatures to different UEs and BSs. In the asymptotic regime, we can show using similar steps as those in~\cite[Proposition~1]{shokri2016Spectrum} that the following holds:
\begin{remark}\label{remark: asymptotic-no-int}
Suppose that a \ac{BS} has perfect CSI toward its associated \acp{UE}.
The interference components, formulated in~\eqref{eq: intra-cell-interference-digital-precoder}--\eqref{eq: inter-operator-interference-digital-precoder}, vanish almost surely as either $N_{\BS} \to \infty$ or $N_{\UE} \to \infty$.
\end{remark}

Remark~\ref{remark: asymptotic-no-int} suggests that we can ignore the intra- and inter-operator coordination completely, and consequently $\calP_1$ and its distributed variant (introduced in Section~\ref{sec: optimal-sharing-strategy}) yield the same optimal solution. Table~\ref{table: Example1} confirms the same trend in the finite antenna regime, where increasing the number of antennas reduces the contributions of coordination on reducing the interference components. Notice that in reality, the perfect CSI assumption of Remark~\ref{remark: asymptotic-no-int} may not hold, leading to a residual sporadic strong interference~\cite{Shokri2015Transitional}. Consequently, we need some level of coordination to tame strong interference terms. However, this mandatory level of coordination at the mmWave bands is much less than that at the sub-6~GHz bands.


\subsection{Imperfect CSI and Hardware}
Although this work alleviates the need for a complete CSI knowledge of the entire network, through the learning functionality, the BSs should have access to error-free effective channels of some selected UEs. However, CSI is estimated using pilots and will inevitably have some estimation errors. These effects are also compounded by the limited number of RF chains in mmWave MIMO, and quantized analog precoding/combining.
But there have been great strides in efficient methods for channel estimation (exploiting sparsity \cite{Alkhateeb2014MIMO} or reciprocity \cite{Ghauch_SED_16}), and hybrid precoding that closely approximates fully digital solutions \cite{Yu_HybridAM_16}.
Moreover, in a distributed setting, CSI acquisition (at the network level) may be done using so-called Forward-Backward training methods to estimate the CSI in a fully distributed manner~\cite{Ghauch_SRMax_17}. These methods, however, may further increase the coordination cost. Sensitivity analysis of the proposed hybrid scheme to the estimation error in the effective channels, convergence with feedback quantization~\cite{magnusson2019maintaining}, and the extension of our approach toward robust learning are important future directions.

\subsection{Signaling and Computational Overheads}
In our approach, we have two sources of signaling. In every CI, we need to acquire CSI from every BS $b$ to UE $u$ for which $[\bC]_{bu} = 1$, whereas the Oracle need CSI for each BS-UE pair. This significantly fewer number of pilot transmissions is feasible due to our rate approximation. To enable it, the cloud collects the current rate measurements from all BSs, re-executes the BCD solver, and announces the new association and coordination (only if they have been changed). This process should be done twice for every training frame, once before the training frame and once after it. Therefore, besides some CSI estimation in every CI, the signaling/communication overhead of the proposed hybrid scheme is mainly dominated by the number of training frames. The frequency of these frames is inevitably large in the first few \acp{CI} since we assume no a priori knowledge about the network. However, we can gradually decrease the exploration frequency by replacing several training frames with operation frames. The lower bound on the exploration frequency depends on many factors, including the dynamics of the topology and the fluctuations of the network load, whose characterization is an interesting topic for future works.

As for the computational complexity, the main contributing factor is solving the two subproblems using BCD (see Algorithm \ref{alg: Cloud}). Although this entails solving two combinatorial problems, one can develop low-complexity solutions, e.g., via relaxations or decompositions. Moreover, the BCD solution is carried out at the cloud server which has large computational resources. Another contributing factor is the matrix inversion in the computation of the RZF precoder at each BS, which scales cubically with the number of UEs served by the BS.

\subsection{Dynamic Number of BSs/UEs}\label{sec: dunamic-no-UEs}
Our main algorithms have been developed for a fixed number of BSs and UEs. In a real network, however, some UEs may join and leave the network, and some BSs may be turned on or off to save energy.

We should highlight that the special characteristics of mmWave communications (directionality, blockage, and propagation loss) would substantially reduce the impact of farther BSs/UEs~\cite{Shokri2018IMSindex}.
In other words, adding/removing some BSs or UEs will have only local effects, impacting the rate models of only a few surrounding UEs. In this situation, the \call{Initialize} function can enable fast adaption to dynamic $\calU$ and $\calB$ using a few new samples.
In the case of having new UEs, we use the \call{Initialize} function for both finding a good initialization for the rate function of the new UEs and for adding some interference terms to the rate models of the existing UEs. In the case of smaller $\calU$, we can remove their impacts on other UEs by removing their contributions to the rate function, approximated by the \call{Initialize} function.

In the light of the above discussion, we argue that the complexity of the functional approximator (e.g., deep neural network) should be manageable in a real network.
The main reason is that the cloud server trains an individual approximator for every UE. In our experiments, our neural network was already over-parameterized. Such a network can easily approximate more complicated rate functions, which may happen for larger $|\calU|$ and $|\calB|$, as we have shown in our experiments over a much bigger network; see Figs.~6 and~7 of the extended version~\cite{shokri2019learhningSPECSreport}. Moreover, due to the interference locality at the mmWave networks~\cite{Shokri2018IMSindex}, a reasonable change in the number of UEs or BSs does not substantially change the hardness of the rate function (to be approximated).
Finally, we reemphasize the fact that current work is intended as a proof of concept of usefulness and viability of the proposed hybrid approach. Several of the issues raised by the reviewers (e.g., scalability and complexity reduction) will be part of our future research.


\section{Conclusions}\label{sec: concluding-remarks}
In this work, we investigated the problem of spectrum sharing in mmWave networks and argued the formidable complexity of a pure model-based solution approach. As a viable alternative, we proposed to complement it by a data-driven approach to make the spectrum sharing problem solvable in practical systems. In particular, the model-based part chooses the beamforming and optimizes association and coordination decisions, given a set of rate models. The data-driven part continuously refines the rate models, maintaining the optimality of our solution even in non-stationary environments. The resulting algorithm balances the use of training frames (designed to explore the solution space) and operation frames (designed to exploit good solutions). Our hybrid scheme has the same computational complexity as the pure model-based approach while being robust to insufficient signaling and missing \ac{CSI}. Our numerical results revealed large gains in network sum-rate while satisfying a predetermined budget on the coordination cost.

\appendices

\section{Proofs}\label{sec: proofs}

\subsection{Lemma~\ref{lemma: BCD-convergence}}
Our assumption that $\widehat{f}_z$ is bi-concave implies that $\widehat{f}_z (\bA, \bC)$ that $\widehat{f}_z$ is concave in $\bA$  for $\bA \in \mathbb{R}^{ |\calB| \times |\calU| } $ when $\bC$ is fixed (and vice versa).
We first show the following inequality holds.
\begin{equation*}
\widehat{f}_z (\bA^{(k)},\bC^{(k)}) \! \overset{(a.1)}{\leq} \! \widehat{f}_z (\bA^{(k+1)},\bC^{(k)}) \! \overset{(a.2)}{\leq}\!  \widehat{f}_z (\bA^{(k+1)},\bC^{(k+1)})
\end{equation*}
Note that $(a.1)$ follows from $\widehat{f}_z$ being concave in $\bA$, which implies that the A-step in \eqref{eq: A-step} has a unique maximizer. Moreover, that maximizer is found due to the exhaustive search solution. Thus, the A-step update cannot decrease $\widehat{f}_z$.
In addition, the same argument can be used to show $(a.2)$: $\widehat{f}_z$ is concave in $\bC$ (meaning that the C-step in \eqref{eq: C-step} has a unique maximizer), and that optimal solution is found (via exhaustive search). Combining $(a.1)$, $(a.2)$, and that $\widehat{f}_z$ is continuous in $(\bA,\bC)$ and bounded above imply that the sequence $\{  \widehat{f}_z (\bA^{(k)},\bC^{(k)})\}_k $ converges to a limit point.

\subsection{Proposition~\ref{prop: initialization}}
When $[\bC]_{iu} = 1$, BS $i$ estimates the effective channel toward UE $u$, namely $\left(\bw_{u}^{\UE}\right)^{\herm}\bH_{iu}$, and uses \ac{RZF} precoder that cancels the interference. Now, let $[\bC]_{iu}=0$ and $[\bA]_{bu}=1$. Assume that we have only LoS links, so $N_{bu} = N_{iu} = 1$, and that $\theta_{b u}^{\UE}$ and $\theta_{i u}^{\UE}$ are AoAs of the LoS links between UE $u$ and BSs $b$ and $i$, respectively. Note that $\theta_{b u}^{\UE}$ and $\theta_{i u}^{\UE}$ can be obtained by the topological knowledge. Define $\phi_{bui} := \theta_{b u}^{\UE} - \theta_{i u}^{\UE}$. Recall the channel model \eqref{eq: channel-matrix} and beamforming models \eqref{eq: AnalogCombiner}--\eqref{eq: DigitalBeamforming}. The interference from unintended BS $i (\neq b)$ is
\begin{subequations}
\begin{align}
& \frac{\sum\limits_{j\in\mathcal{A}_{i}}\mathbb{E}\left[\lambda_{i}\left| \left(\bw_{u}^{\UE}\right)^{\herm} \bH_{iu} \bw_{i j}^{\BS} \right|^2 \right]}{N_{\BS}N_{\UE}} \\
& \hspace{1mm} \stackrel{\text{\eqref{eq: channel-matrix}}}{=} \sum\limits_{j\in\mathcal{A}_{i}}\mathbb{E}\left[\lambda_{i}|g_{i u}|^2 \left| \left(\bw_{u}^{\UE}\right)^{\herm}  \ba_{\UE}\left(\theta_{i u}^{\UE}\right) \ba_{\BS}^{\herm} \left(\theta_{i u}^{\BS}\right)  \bw_{i j}^{\BS} \right|^2 \right]  \\
& \hspace{1mm} \stackrel{\text{\eqref{eq: AnalogCombiner}}}{=} \sum\limits_{j\in\mathcal{A}_{i}}\mathbb{E}\left[\lambda_{i}|g_{i u}|^2 \left| \ba_{\UE}^{\herm}\left(\theta_{b u}^{\UE}\right)  \ba_{\UE}\left(\theta_{i u}^{\UE}\right) \ba_{\BS}^{\herm} \left(\theta_{i u}^{\BS}\right)  \bw_{i j}^{\BS} \right|^2 \right]  \\
& \hspace{1mm} \stackrel{\text{(a)}}{=} L_{iu} \left|\frac{\sin\left(\frac{N_{\UE} \phi_{bui}}{2}\right)}{\frac{N_{\UE} \phi_{bui}}{2}}\right| \sum\limits_{j\in\mathcal{A}_{i}}\mathbb{E}\left[\lambda_{i} \left| \ba_{\BS}^{\herm} \left(\theta_{i u}^{\BS}\right)  \bw_{i j}^{\BS} \right|^2 \right]
\\
& \hspace{1mm} \stackrel{\text{(b)}}{=} L_{iu} \left|\sinc\left(\frac{N_{\UE} (\theta_{b u}^{\UE} - \theta_{i u}^{\UE})}{2}\right)\right| \rho^{\mathrm{Tx}} \:,
\end{align}
\end{subequations}
where the expectations are over the randomness on the channel gains and consequently on the beamforming vectors, (a) is due to the mutual independence of $\ba_{\UE}(\theta_{i u}^{\UE})$, $\ba_{\BS} (\theta_{i u}^{\BS})$, and $\bw_{i j}^{\BS}$ when $b\neq i$, and (b) is due to \eqref{eq: power-normalization-precoder} and that $\ba_{\BS}^{\herm} (\theta_{i u}^{\BS})\ba_{\BS} (\theta_{i u}^{\BS}) = 1$.


\bibliographystyle{IEEEtran}
\bibliography{References,Ref2}

\begin{thebibliography}{10}
\providecommand{\url}[1]{#1}
\csname url@samestyle\endcsname
\providecommand{\newblock}{\relax}
\providecommand{\bibinfo}[2]{#2}
\providecommand{\BIBentrySTDinterwordspacing}{\spaceskip=0pt\relax}
\providecommand{\BIBentryALTinterwordstretchfactor}{4}
\providecommand{\BIBentryALTinterwordspacing}{\spaceskip=\fontdimen2\font plus
\BIBentryALTinterwordstretchfactor\fontdimen3\font minus
  \fontdimen4\font\relax}
\providecommand{\BIBforeignlanguage}[2]{{%
\expandafter\ifx\csname l@#1\endcsname\relax
\typeout{** WARNING: IEEEtran.bst: No hyphenation pattern has been}%
\typeout{** loaded for the language `#1'. Using the pattern for}%
\typeout{** the default language instead.}%
\else
\language=\csname l@#1\endcsname
\fi
#2}}
\providecommand{\BIBdecl}{\relax}
\BIBdecl

\bibitem{Jiang2018LowLatency}
X.~Jiang, H.~S. Ghadikolaei, G.~Fodor, E.~Modiano, Z.~Pang, M.~Zorzi, and
  C.~Fischione, ``Low-latency networking: {W}here latency lurks and how to tame
  it,'' \emph{Proc. IEEE}, vol. 107, no.~2, pp. 280--306, Feb. 2019.

\bibitem{FCC:2016}
{47 CFR Parts 2, 25, 30}, \emph{Use of Spectrum Bands Above 24 GHz for Mobile
  Radio Services; Proposed Rule}, Federal Register Std. Vol. 81, No. 164, Part
  IV, August 2016.

\bibitem{Boccardi:16}
F.~Boccardi, H.~S. Ghadikolaei, G.~Fodor, E.~Erkip, C.~Fischione,
  M.~Kountouris, P.~Popovski, and M.~Zorzi, ``Spectrum pooling in mmwave
  networks: {O}pportunities, challenges, and enablers,'' \emph{IEEE Commun.
  Mag.}, pp. 33--39, November 2016.

\bibitem{Rebato:17}
M.~Rebato, F.~Boccardi, M.~Mezzavilla, S.~Rangan, and M.~Zorzi, ``Hybrid
  specturm sharing in mmwave cellular networks,'' \emph{{IEEE} Trans. Cogn.
  Commun. Netw.}, vol.~3, no.~2, pp. 155--168, June 2017.

\bibitem{Hu:18}
F.~Hu, B.~Chen, and K.~Zhu, ``Full spectrum sharing in cognitive radio networks
  toward {5G}: {A} survey,'' \emph{IEEE Access}, vol.~6, pp. 15\,754--5776,
  April 2018.

\bibitem{Jurdi:18}
R.~Jurdi, A.~K. Gupta, J.~G. Andrews, and R.~W. Heath, ``Modeling
  infrastructure sharing in mmwave networks with shared spectrum licenses,''
  \emph{IEEE Trans. Cognitive Comm. and Networking}, pp. 1--18, March 2018.

\bibitem{Doyle:14}
L.~Doyle, J.~Kibilda, T.~K. Forde, and L.~DaSilva, ``Spectrum without bounds,
  networks without borders,'' \emph{Proc. IEEE}, vol. 102, no.~3, pp. 351--365,
  March 2014.

\bibitem{shokri2016Spectrum}
H.~S.~Ghadikolaei \emph{et~al.}, ``Spectrum sharing in {mmWave} cellular
  networks via cell association, coordination, and beamforming,'' \emph{{IEEE}
  J. Sel. Areas Commun.}, vol.~34, no.~11, pp. 2902--2917, Nov. 2016.

\bibitem{Tsiftsis:16}
T.~A. Tsiftsis, G.~Ding, Y.~Zou, G.~K. Karagiannidis, Z.~Han, and L.~Hanzo,
  ``Guest editorial: Spectrum sharing and aggregation for future wireless
  networks, part i,'' \emph{{IEEE} J. Sel. Areas Commun.}, vol.~34, no.~10, pp.
  2533--2536, Oct. 2016.

\bibitem{Mihovska:09}
A.~Mihovska \emph{et~al.}, ``Multi-operator resource sharing scenario in the
  context of {IMT}-advanced systems,'' in \emph{Second International Workshop
  on Cognitive Radio and Advanced Spectrum Management}, Aalborg, Denmark, May
  2009.

\bibitem{McMenamy:14}
J.~McMenamy, I.~Macaluso, N.~Marchetti, and L.~Doyle, ``A methodology to help
  operators share the spectrum through an enhanced form of carrier
  aggregation,'' in \emph{IEEE International Symposium on Dynamic Spectrum
  Access Networks (DYSPAN)}, McLean, VA, USA, 1-4 April 2014, pp. 334--344.

\bibitem{Holland:15}
O.~Holland and M.~Dohler, ``Geolocation-based architecture for heterogeneous
  spectrum usage in {5G},'' in \emph{IEEE Globecom Workshops}, San Diego, CA,
  USA, Dec. 2015, pp. 1--6.

\bibitem{Krysz:16}
P.~Kryszkiewicz, A.~Kliks, and H.~Bogucka, ``Small-scale spectrum aggregation
  and sharing,'' \emph{{IEEE} J. Sel. Areas Commun.}, vol.~34, no.~10, pp.
  2630--2641, October 2016.

\bibitem{Xiao:16}
Y.~Xiao, Z.~Han, C.~Yuen, and L.~A. DaSilva, ``Carrier aggregation between
  operators in next generation cellular networks: A stable roommate market,''
  \emph{IEEE Trans. Wireless Comm.}, vol.~15, no.~1, pp. 633--649, Jan. 2016.

\bibitem{Hossain:14}
C.~{Clancy}, J.~{Hecker}, E.~{Stuntebeck}, and T.~{O'Shea}, ``Applications of
  machine learning to cognitive radio networks,'' \emph{IEEE Wireless
  Communications}, vol.~14, no.~4, pp. 47--52, August 2007.

\bibitem{Zhang:15}
Z.~Zhang, K.~Zhang, F.~Gao, and S.~Zhang, ``Spectrum prediction and channel
  selection for sensing-based spectrum sharing scheme using online learning
  techniques,'' in \emph{Proc. IEEE PIMRC}, 2015.

\bibitem{Srinivasan:16}
M.~Srinivasan, V.~J. Kotagi, and C.~S.~R. Murthy, ``A {Q}-learning framework
  for user {QoE} enhanced self-organizing spectrally efficient network using a
  novel inter-operator proximal spectrum sharing,'' \emph{{IEEE} J. Sel. Areas
  Commun.}, vol.~34, no.~11, pp. 2887--2901, Nov. 2016.

\bibitem{tse2005fundamentals}
D.~Tse and P.~Viswanath, \emph{Fundamentals of wireless communication}.\hskip
  1em plus 0.5em minus 0.4em\relax Cambridge University Press, 2005.

\bibitem{Shokri2018IMSindex}
H.~S. Ghadikolaei, C.~Fischione, and E.~Modiano, ``Interference model
  similarity index and its applications to {mmWave} networks,'' \emph{{IEEE}
  Trans. Wireless Commun}, vol.~17, no.~1, pp. 71--85, Jan. 2018.

\bibitem{Sevakula:15}
R.~K. Sevakula, M.~Suhail, and N.~K. Verma, ``Fast data sampling for large
  scale support vector machines,'' in \emph{IEEE Workshop on Computational
  Intelligence: Theories, Applications and Future Directions}, Kanpur, India,
  Dec. 2015.

\bibitem{gai2012combinatorial}
Y.~Gai, B.~Krishnamachari, and R.~Jain, ``Combinatorial network optimization
  with unknown variables: {Multi-armed} bandits with linear rewards and
  individual observations,'' \emph{IEEE/ACM Trans. Netw.}, vol.~20, no.~5, pp.
  1466--1478, Oct. 2012.

\bibitem{laufer2018hybrid}
B.~Laufer-Goldshtein, R.~Talmon, and S.~Gannot, ``A hybrid approach for speaker
  tracking based on {TDOA} and data-driven models,'' \emph{IEEE/ACM Trans.
  Audio, Speech, Language Process.}, vol.~26, no.~4, pp. 725--735, Apr. 2018.

\bibitem{zappone2018model}
A.~{Zappone}, M.~{Di Renzo}, M.~{Debbah}, T.~T. {Lam}, and X.~{Qian},
  ``Model-aided wireless artificial intelligence: Embedding expert knowledge in
  deep neural networks for wireless system optimization,'' \emph{IEEE Vehicular
  Technology Magazine}, vol.~14, no.~3, pp. 60--69, Sep. 2019.

\bibitem{zappone2019wireless}
A.~{Zappone}, M.~{Di Renzo}, and M.~{Debbah}, ``Wireless networks design in the
  era of deep learning: Model-based, ai-based, or both?'' \emph{IEEE
  Transactions on Communications}, vol.~67, no.~10, pp. 7331--7376, Oct 2019.

\bibitem{shokri2019learhningSPECSreport}
H.~S. Ghadikolaei, H.~Ghauch, G.~Fodor, M.~Skoglund, and C.~Fischione, ``A
  hybrid model-based and data-driven approach to spectrum sharing in {mmWave}
  cellular networks: {E}xtended version,'' \emph{arXiv preprint
  arXiv:1412.6980}, 2020.

\bibitem{andrews2014overview}
J.~G. Andrews, S.~Singh, Q.~Ye, X.~Lin, and H.~S. Dhillon, ``An overview of
  load balancing in {HetNets}: {O}ld myths and open problems,'' \emph{{IEEE}
  Wireless Commun.}, vol.~21, no.~2, pp. 18--25, Apr. 2014.

\bibitem{Akdeniz2014MillimeterWave}
M.~Akdeniz, Y.~Liu, M.~Samimi, S.~Sun, S.~Rangan, T.~Rappaport, and E.~Erkip,
  ``Millimeter wave channel modeling and cellular capacity evaluation,''
  \emph{{IEEE} J. Sel. Areas Commun.}, vol.~32, no.~6, pp. 1164--1179, Jun.
  2014.

\bibitem{Ayach2012Capacity}
O.~Ayach \emph{et~al.}, ``The capacity optimality of beam steering in large
  millimeter wave {MIMO} systems,'' in \emph{Proc. IEEE International Workshop
  on Signal Processing Advances in Wireless Communications}, 2012, pp.
  100--104.

\bibitem{Yu_HybridAM_16}
X.~Yu, J.~Shen, J.~Zhang, and K.~B. Letaief, ``Alternating minimization
  algorithms for hybrid precoding in millimeter wave {MIMO} systems,''
  \emph{IEEE Journal of Selected Topics in Signal Processing}, vol.~10, no.~3,
  pp. 485--500, Apr. 2016.

\bibitem{Ghauch_SED_16}
H.~Ghauch, T.~Kim, M.~Bengtsson, and M.~Skoglund, ``Subspace estimation and
  decomposition for large millimeter-wave {MIMO} systems,'' \emph{IEEE Journal
  of Selected Topics in Signal Processing}, vol.~10, no.~3, pp. 528--542, Apr.
  2016.

\bibitem{di2014stochastic}
M.~Di~Renzo, ``Stochastic geometry modeling and analysis of multi-tier
  millimeter wave cellular networks,'' \emph{{IEEE} Trans. Wireless Commun.},
  vol.~14, no.~9, pp. 5038--5057, Sept. 2015.

\bibitem{Yu2016Distributed}
Y.~Xu, H.~S. Ghadikolaei, and C.~Fischione, ``Distributed association and
  relaying with fairness in millimeterwaves networks,'' \emph{{IEEE} Trans.
  Wireless Commun.}, vol.~15, no.~12, pp. 7955--7970, Dec. 2016.

\bibitem{Park2009Analysis}
M.~Park and P.~Gopalakrishnan, ``Analysis on spatial reuse and interference in
  60-{GHz} wireless networks,'' \emph{{IEEE} J. Sel. Areas Commun.}, vol.~27,
  no.~8, pp. 1443--1452, Oct. 2009.

\bibitem{Razaviyayn_BCD_12}
M.~Razaviyayn, M.~Hong, and Z.-Q. Luo, ``A unified convergence analysis of
  block successive minimization methods for nonsmooth optimization,''
  \emph{SIAM Journal on Optimization}, vol.~23, no.~2, pp. 1126--1153, June
  2013.

\bibitem{Berstakas_nonlinear_99}
D.~Bertsekas, \emph{Nonlinear Programming}, 2nd~ed.\hskip 1em plus 0.5em minus
  0.4em\relax Athena Scientific, 1999.

\bibitem{jin2017escape}
C.~Jin, R.~Ge, P.~Netrapalli, S.~M. Kakade, and M.~I. Jordan, ``How to escape
  saddle points efficiently,'' in \emph{Proc. International Conference on
  Machine Learning (ICML)}.\hskip 1em plus 0.5em minus 0.4em\relax JMLR. org,
  2017, pp. 1724--1732.

\bibitem{Shokri2015Transitional}
H.~S. Ghadikolaei and C.~Fischione, ``The transitional behavior of interference
  in millimeter wave networks and its impact on medium access control,''
  \emph{{IEEE} Trans. Commun.}, vol.~62, no.~2, pp. 723--740, Feb. 2016.

\bibitem{verdu1998multiuser}
S.~Verdu, \emph{Multiuser detection}.\hskip 1em plus 0.5em minus 0.4em\relax
  Cambridge university press, 1998.

\bibitem{Athanasiou-etal-2013}
G.~Athanasiou, C.~Weeraddana, C.~Fischione, and L.~Tassiulas, ``Optimizing
  client association in {60 GHz} wireless access networks,'' \emph{{IEEE/ACM}
  Trans. Netw.}, vol.~23, no.~3, pp. 836--850, Jun. 2015.

\bibitem{Bottou2018SIAM}
L.~Bottou, F.~Curtis, and J.~Nocedal, ``Optimization methods for large-scale
  machine learning,'' \emph{SIAM Review}, vol.~60, no.~2, pp. 223--311, 2018.

\bibitem{kingma2014adam}
D.~P. Kingma and J.~Ba, ``{ADAM}: {A} method for stochastic optimization,'' in
  \emph{Proc. International Conference on Learning Representations (ICLR)},
  2015.

\bibitem{Alkhateeb2014MIMO}
A.~Alkhateeb, J.~Mo, N.~González-Prelcic, and R.~Heath, ``{MIMO} precoding and
  combining solutions for millimeter-wave systems,'' \emph{{IEEE} Commun.
  Mag.}, vol.~52, no.~12, pp. 122--130, Dec. 2014.

\bibitem{Ghauch_SRMax_17}
H.~Ghauch, T.~Kim, M.~Bengtsson, and M.~Skoglund, ``Sum-rate maximization in
  sub-28-{GHz} millimeter-wave {MIMO} interfering networks,'' \emph{{IEEE} J.
  Sel. Areas Commun.}, vol.~35, no.~7, pp. 1649--1662, Jul. 2017.

\bibitem{magnusson2019maintaining}
S.~Magn{\'u}sson, H.~S.~Ghadikolaei, and N.~Li, ``On maintaining linear
  convergence of distributed learning and optimization under limited
  communication,'' \emph{arXiv preprint arXiv:1902.11163}, 2019.

\end{thebibliography}

\end{document}